\documentclass[pra,twocolumn,floatfix,showpacs]{revtex4}
\usepackage{amsmath,amsfonts}
\usepackage{graphicx}

\renewcommand{\vec}{\boldsymbol}
\newcommand{\re}{e}
\newcommand{\ri}{i}

\begin{document}

\title{Semiclassical quantization of the hydrogen atom in crossed electric
       and magnetic fields}
\author{Thomas Bartsch}
\author{J\"org Main}
\author{G\"unter Wunner}
\affiliation{Institut f\"ur Theoretische Physik 1, Universit\"at Stuttgart,
         D-70550 Stuttgart, Germany}
\date{\today}

\begin{abstract}
The $S$-matrix theory formulation of closed-orbit theory recently proposed
by Granger and Greene is extended to atoms in crossed
electric and magnetic fields. We then present a  semiclassical
quantization of the hydrogen atom in crossed fields, which succeeds in
resolving individual lines in the spectrum, but is restricted to the
strongest lines of each $n$-manifold. By means of a detailed semiclassical
analysis of the quantum spectrum, we demonstrate that it is the abundance
of bifurcations of closed orbits that precludes the resolution of finer
details. They necessitate the inclusion of uniform semiclassical
approximations into the quantization process. Uniform approximations for
the generic types of closed-orbit bifurcation are derived, and a general
method for including them in a high-resolution semiclassical quantization is
devised.
\end{abstract}
\maketitle

\section{Introduction}
\label{sec:Intro}

Closed-orbit theory was first introduced by Du and Delos~\cite{Du88} and
Bogomolny~\cite{Bogomolny89} some twenty years ago to interpret the
modulations observed in the photo-absorp\-tion spectra of hydrogenic Rydberg
atoms in a magnetic field close to the ionization threshold. Since that
time, it turned out to be a powerful and flexible tool for the semiclassical
interpretation of a variety of spectra.
It has been used to describe atoms
in electric~\cite{Gao92} as well as parallel~\cite{Mao93,Courtney95a} or
crossed~\cite{Main91,Raithel91,Peters93} electric and magnetic
fields. In the case of non-hydrogenic atoms, the influence of the ionic
core can be modelled either by means of an effective classical
potential~\cite{Huepper96,Weibert98} or in terms of quantum
defects~\cite{Dando95,Dando96}. Recently, closed-orbit theory has even been
shown to be applicable to the spectra of simple molecules in external
fields~\cite{Matzkin01}.

A complete description of photo-absorption spectra requires the calculation
of the energies $E_n$ of the excited atomic states and the strengths of the
spectral lines, which is characterized by the dipole matrix elements
$\left<i|D|n\right>$ between the initial state $\left|i\right>$ and the
Rydberg state $\left|n\right>$, where $D$ is the component of the dipole
operator describing the polarization of the exciting laser field. These
quantities are neatly summarized in the response function
\begin{equation}
  \label{gDef}
  g(E)=-\frac 1{\pi}\left<i|D\,G(E)\,D|i\right> 
      =-\frac 1{\pi}
       \sum_n \frac{|\left<i|D|n\right>|^2}{E-E_n+\ri\epsilon} \;,
\end{equation}
where 
\begin{equation}
  G(E) = \sum_n \frac{|n\rangle\langle n|}{E-E_n+\ri\epsilon}
\end{equation}
denotes the retarded Green's function.

Closed-orbit theory provides a semiclassical approximation to the quantum
response function~(\ref{gDef}), which splits into a smooth part and an
oscillatory part of the form
\begin{equation}
  \label{resOscGen}
  g^{\rm osc}(E) =
    \sum_{\rm c.o.} {\cal A}_{\rm c.o.}(E)\, \re^{\ri S_{\rm c.o.}(E)} \;,
\end{equation}
where the sum extends over all classical closed orbits starting from the
nucleus and returning to it after having been deflected by the external
fields, $S_{\rm c.o.}$ is the classical action of the closed orbit, and the
amplitude $\cal A_{\rm c.o.}$ describes its stability and its starting and
returning directions. Its precise form depends on the geometry of the
external fields. In section~\ref{sec:Smatrix}, it will be specified for
systems with and without a rotational symmetry.

Although the closed-orbit sum~(\ref{resOscGen}) appears to provide a
straightforward means of calculating the response function from the
classical closed orbits, this is actually not the case because the sum
usually diverges for real energies $E$. Thus, the quantal information
cannot be extracted directly from the semiclassical expansion. One
particular and widely applicable method to overcome the convergence
problems of the closed-orbit sum is semiclassical quantization by harmonic
inversion \cite{Main98b,Main99d}. For the hydrogen atom in a
magnetic field, this method has been shown \cite{Main99a} to be capable of
extracting semiclassical eigenenergies and transition matrix elements from
a closed-orbit sum.

In the present paper we will investigate how these results can be
generalized to the hydrogen atom in crossed electric and magnetic
fields. This problem is considerably harder than the treatment of the
diamagnetic hydrogen atom, which possesses a rotational symmetry around the
field axis. Due to that symmetry, in classical mechanics the angular
momentum around the field axis is conserved. So is, in quantum mechanics,
the magnetic quantum number $m$.  In crossed fields, the rotational
symmetry is broken. As a consequence, the selection rules for the
$m$-quantum number no longer hold, and a multitude of additional lines
appears in the quantum spectrum. At the same time, the determination of
classical closed orbits gets significantly more difficult because three
non-separable degrees of freedom have to be dealt with.  A detailed
description of the intricate pattern of closed orbits and their
bifurcations was given in an accompanying paper \cite{Bartsch03a}. That
data forms the basis of the present work, where the semiclassical treatment
of the crossed-fields hydrogen atom will be dealt with, and we will freely
use the nomenclature introduced in \cite{Bartsch03a}.

After the essential properties of the crossed-fields Hamiltonian have been
summarized in section~\ref{sec:classHam}, we start, in
section~\ref{sec:Smatrix}, with a derivation of the closed-orbit
formula~(\ref{resOscGen}) in the context of the $S$-matrix formulation of
closed-orbit theory introduced recently by Granger and Greene
\cite{Granger00}. We show that the novel framework can be extended to the
crossed-fields situation, and we clarify some misleading conclusions
arrived at in \cite{Granger00}. Section~\ref{sec:Xqm} describes the quantum
spectrum under study, and section~\ref{sec:SclLo} compares it to a
semiclassical spectrum in low resolution. In section~\ref{sec:SclHi}, the
results of a high-resolution semiclassical quantization using the technique
of harmonic inversion are presented. The semiclassical spectrum correctly
identifies the strongest spectral lines, but it fails to describe finer
details of the quantum spectrum.  In section~\ref{sec:SclRec}, we compare a
quantum recurrence spectrum to the classical data to show that the
principal source of this difficulty lies in the abundance of closed-orbit
bifurcations. Uniform approximations provide a tool to cope with the
divergences introduced into semiclassical spectra by bifurcations of
classical orbits. A general technique for their construction is described
in section~\ref{sec:Uniform}, and uniform approximations for the two types
of generic codimension-one bifurcations identified in \cite{Bartsch03a} are
derived. Finally, in section~\ref{ssec:UnifRecur} we demonstrate how
uniform approximations can be incorporated into recurrence spectra, thus
paving the way for their inclusion into the high-resolution semiclassical
quantization by harmonic inversion.

\section{The classical Hamiltonian}
\label{sec:classHam}

Throughout this work, we will assume the magnetic field to be directed
along the $z$-axis and the electric field to be directed along the
$x$-axis. In atomic units, the Hamiltonian describing the motion of the
atomic electron then reads
\begin{equation}
  \label{specHam}
  H=\frac12\vec p^2-\frac 1r + \frac 12BL_z+\frac 18B^2\rho^2+Fx \;,
\end{equation}
where $B$ and $F$ denote the magnetic and electric field strengths,
respectively, $r^2=x^2+y^2+z^2$, $\rho^2=x^2+y^2$, and $L_z$ is the
$z$-component of the angular momentum vector.  By virtue of the scaling
properties of the Hamiltonian~(\ref{specHam}), if all classical quantities
are multiplied by suitable powers of the scaling parameter 
\begin{equation}
  w\equiv B^{-1/3}
\end{equation}
the dynamics can be shown not to depend on the energy $E$ and the field
strengths $B$ and $F$ separately, but only on the scaled energy $\tilde
E=w^2E$ and the scaled electric field strength $\tilde F=w^4F$. In
particular, classical actions scale according to $S=w\tilde S$. Thus, the
semiclassical limit of large classical actions corresponds to the limit of
large $w$.

The way of recording a quantum spectrum which is best suited for
semiclassical investigations is scaled-energy spectroscopy. A spectrum then
consists of a list of the scaling parameters $w_n$ characterizing the
quantum states for given scaled energy $\tilde E$ and scaled electric
field strength $\tilde F$. Scaled-energy spectroscopy offers the advantage
that the underlying classical dynamics does not change across the
spectrum. It will be adopted throughout this work.

\section{The $S$-matrix formulation of closed-orbit theory}
\label{sec:Smatrix}

\subsection{General formalism}
\label{ssec:SGen}

The basic observation fundamental to all of closed-orbit theory is a
partition of space into physically distinct regions. In the core region
close to the nucleus, the Rydberg electron interacts in a complicated
manner with all electrons of the ionic core. This interaction is manifestly
quantum mechanical in nature, it cannot be described in the framework of
semiclassical theories. On the other hand, the interaction of the Rydberg
electron with the external fields is much weaker in the core region than
its interaction with the core, so that the fields can safely be
neglected. Therefore, a description of the core obtained in the field-free
case can be used. In particular, the initial state of the photo-absorption
process is assumed to be localized in the core region and not to be
influenced by the external fields.

In the long-range region far away from the nucleus, on the other hand, the
external fields play a dominant role, whereas there is no interaction with
the ionic core except for the Coulomb attraction of its residual charge. In
this region, the dynamics of the Rydberg electron is well-suited for a
semiclassical description. It is independent of the details of the ionic
core.

In order to establish a link between the dynamics in the core and
long-range regions, a matching region is assumed to exist at
intermediate distances from the nucleus where both the external fields and
the interaction with the core are negligible. Thus, in the
matching region the simple physics of an electron subject to the
residual Coulomb field of the core is observed.

Recently, Granger and Greene~\cite{Granger00} proposed a novel formulation
of the theory based on ideas borrowed from quantum-defect theory. Their
formulation achieves a clear separation between properties of the external
field configuration and the ionic core, which are encoded in separate
$S$-matrices. Suitable approximations to the core and the long-range
$S$-matrices can be derived independently. Therefore, the formalism can be
expected to allow a generalization of closed-orbit theory to atoms with
ionic cores exhibiting more complicated internal dynamics than have been
treated so far.

The derivation given by Granger and Greene treated the case of an atom in a
magnetic field only. It will now be extended in such a way that it holds
for combined electric and magnetic fields with arbitrary field
configurations. To this end, the ansatz and basic formulae of  Granger and
Greene's theory will be summarized in this section. A more detailed
treatment can be found in their paper~\cite{Granger00}. In subsequent
sections, we will then turn to a discussion of the long-range scattering
matrices pertinent to different external field configurations.

To lay the foundation for a definition of the $S$-matrices, we pick a basis
set $\Psi_k^{\rm core}$ and $\Psi_k^{\rm LR}$ of wave functions of the
Rydberg electron valid in the core and long-range regions, respectively,
and expand in terms of spherical harmonics
\begin{equation}
  \label{PsiS}
  \Psi_k^{\rm core(LR)}(r,\vartheta,\varphi) = 
    \frac 1r \sum_{k'} Y_{k'}(\vartheta,\varphi) F_{k'k}^{\rm core(LR)}(r)
    \;.
\end{equation}
The channel index $k$ is to be understood as a double index $(l,m)$
characterizing the spherical harmonics. When studying a complicated atom
with more than one relevant state of the core, additional information
can be included in the channel functions $Y_k$.

In the matching region, the radial function matrices $\underline{F}^{\rm core}$
and $\underline{F}^{\rm LR}$ can both be expressed in terms of radial Coulomb
functions. We use the functions $f^+_k(r)$ and $f^-_k(r)$ satisfying
outgoing and incoming wave boundary conditions, respectively,
given by~\cite{Aymar96} and choose the radial functions to be of the
form
\footnote{The regular and irregular Coulomb functions $f$ and $g$ used
  by Granger and Greene~\cite{Granger00} differ from those used by
  Robicheaux~\cite{Robicheaux93} in that they are energy-normalized 
  in Rydberg rather than in Hartree units.  The radial function matrices
  given here agree in normalization with those adopted by Robicheaux,
  whereas the matrices used by Granger and Greene are
  inconsistent with their equation (12).}
\begin{equation}
  \label{Fcore}
  F_{k'k}^{\rm core}(r) = -\ri
    \left[f_{k'}^+(r)\,S_{k'k}^{\rm core} - f_{k'}^-(r)\,\delta_{k'k}\right]\;,
\end{equation}
\begin{equation}
  \label{FLR}
  F_{k'k}^{\rm LR}(r) = -\ri
    \left[f_{k'}^+(r)\,\delta_{k'k} - f_{k'}^-(r)\,S_{k'k}^{\rm LR}\right]\;.
\end{equation}
Physically, these choices mean that the basis function $\Psi_k^{\rm core}$
is a superposition of a single incoming wave in channel $k$ and the
outgoing waves in different channels produced from it by scattering off the
core. Similarly, $\Psi_k^{\rm LR}$ consists of an outgoing wave in channel
$k$ and the returning waves generated by scattering off the external
fields. The scattering matrices $\underline{S}^{\rm core}$ and
$\underline{S}^{\rm LR}$ thus summarize the physical properties of the core
and the external fields, respectively. They are determined by the condition
that the radial functions obey suitable boundary conditions,
i.e. $\underline{F}^{\rm core}$ is regular at the origin, whereas
$\underline{F}^{\rm LR}$ vanishes or satisfies outgoing-wave boundary
conditions at infinity for bound and free states, respectively. For
hydrogen, $\underline{S}^{\rm core}$ is the identity matrix.

Following previous work by Robicheaux~\cite{Robicheaux93}, Granger and
Greene derive the following expression for the response 
function~(\ref{gDef}):
\begin{equation}
  \label{resSeries}
  \begin{split}
  g=\ri\,\underline{d}^\dagger
     \Big(\underline 1
         &+2\left(\underline{S}^{\rm core}\underline{S}^{\rm LR}\right)
          +2\left(\underline{S}^{\rm core}\underline{S}^{\rm LR}\right)^2 \\
         &+2\left(\underline{S}^{\rm core}\underline{S}^{\rm LR}\right)^3
          +\dots\Big) \underline{d} \;,
  \end{split}
\end{equation}
where the vector $\underline{d}$ comprises the energy-dependent dipole matrix
elements
\begin{equation}
  \label{channelMat}
  d_k(E) = \left<\Psi_k^{\rm core}(E)|D|i\right>
\end{equation}
between the initial state and the core-region channel wave
functions. For hydrogen they can be computed explicitly (see, e.g.,
\cite{Du88} or \cite{Bartsch02}).

The terms of the series~(\ref{resSeries}) embody contributions from paths
where the Rydberg electron takes zero, one, two, etc.~trips out into the
long-range region and back to the core before interfering with the initial
outgoing wave. In the semiclassical approximation, $\underline{S}^{\rm LR}$
will be given in terms of closed orbits. A returning wave is associated
with each returning classical orbit. By a general ionic core, it is
scattered into all directions. The parts of the wave scattered into the
outgoing direction of a closed orbit will then follow this orbit until they
return to the core again. Thus, core scattering leads to a concatenation of
different closed orbits~\cite{Dando95,Dando96}. In hydrogen, the Coulomb
center scatters the incoming wave back into its direction of incidence, so
that there is no coupling of closed orbits. Terms describing repeated
scattering off the external fields are therefore absent from the sum, and
the hydrogen response function can be decomposed into a smooth part
\begin{equation}
  \label{resSmooth}
  g_0=\ri\,\underline{d}^\dagger \underline{d} \;,
\end{equation}
which is the same as in the field-free case and contains ``direct''
contributions where the electron does not scatter off the external fields
at all, and an oscillatory part
\begin{equation}
  \label{resOsc}
  g^{\rm osc} = 2\ri\,\underline{d}^\dagger \underline{S}^{\rm LR} \underline{d}
\end{equation}
generated by the electron going out into the long-range region and being
scattered back to the nucleus. It is this part which describes the impact
of the external fields.

The basis for a semiclassical approximation is provided by the retarded
Green's function $G(\vec x,\vec x';E)$ describing the propagation of the
electron from $\vec x'$ to $\vec x$ at the energy $E$.  It can be expanded
in terms of the channel functions as
\begin{equation}
  \label{GreenExp}
  G(\vec x,\vec x';E)=\frac 1{rr'}
     \sum_{kk'}Y_k(\vartheta,\varphi)\,\widetilde G_{kk'}(r,r';E)\,
               Y_{k'}^\ast (\vartheta',\varphi')
\end{equation}
with
\begin{equation}
  \label{GreenChannel}
  \widetilde G_{kk'}(r,r';E) = rr'\left<k|G(\vec x,\vec x';E)|k'\right> \;.
\end{equation}
The long-range scattering
matrix is related to the Green's function matrix by \cite{Granger00}
\begin{equation}
  \label{SGreen}
  \underline{S}^{\rm LR} = \frac{1}{\ri\pi} [\underline{f}^-(r_0)]^{-1}
     \underline{G}(r_0,r_0) [\underline{f}^-(r_0)]^{-1} \;,
\end{equation}
where $r_0$ is the matching radius, $\underline f^-$ is the diagonal matrix
\begin{equation}
  f^-_{kk'}(r)=f^-_k(r)\,\delta_{kk'}
\end{equation}
comprising the radial wave functions, and
$\underline{G}(r,r')$ denotes the part of $\widetilde{\underline{G}}(r,r')$ satisfying
incoming-wave boundary conditions at the final radius $r$. The latter
condition ensures that only electron paths approaching the matching radius
from the long-range region contribute to $\underline{S}^{\rm LR}$, whereas paths
that traverse the core region are omitted.

\subsection{Closed-orbit theory for crossed-fields systems}
\label{sec:COCrossed}

To obtain a semiclassical approximation to the long-range scattering
matrix, we make use of the semiclassical Green's function derived by
Gutzwiller~\cite{Gutzwiller90}
\begin{equation}
  \label{Gscl}
  \begin{split}
  G^{\rm scl}(\vec x, \vec x';E) =& \frac{2\pi}{(2\pi\ri)^{(n+1)/2}} \\
    &\times
    \sum_{\rm class.~traj.} \sqrt{|D|}
      \exp\left(\ri S-\ri\frac{\pi}2\sigma\right) \;,
  \end{split}
\end{equation}
where the sum extends over all classical trajectories leading from $\vec
x'$ to $\vec x$ at the energy $E$, $n$ is the number of degrees of freedom,
$S$ is the classical action along the trajectory, $\sigma$ the number of
caustics along the trajectory, and
\begin{equation}
  \label{GreenDDef}
  D=\det\left(
      \begin{array}{cc}
        \frac{\partial^2 S}{\partial\vec x \partial\vec x'} &
        \frac{\partial^2 S}{\partial\vec x \partial E} \\[.5ex]
        \frac{\partial^2 S}{\partial E \partial\vec x'} &
        \frac{\partial^2 S}{\partial E^2}
      \end{array}
    \right)
\end{equation}
is the amplitude for the contribution of the trajectory. By
(\ref{GreenChannel}), we obtain a semiclassical
approximation to the Green's function matrix
\begin{equation}
  \label{SclGreenChannel}
  \begin{split}
  &G^{\rm scl}_{kk'} (r_0,r_0;E) =
    \frac{2\pi}{(2\pi\ri)^2}r_0^2
    \int d\vartheta\,d\vartheta'\,d\varphi\,d\varphi'
         \sin\vartheta\sin\vartheta'\; \\
         &\quad\times 
           Y_k^\ast (\vartheta,\varphi)\, Y_{k'}(\vartheta',\varphi')
           \sum_{\rm class.~traj.} \sqrt{|D|}
             \re^{\ri (S(r_0,r_0)-\pi\sigma/2)}\;.
  \end{split}
\end{equation}
As usual in semiclassics, the integrals will be evaluated in the
stationary-phase approximation. It yields a sum over all classical
trajectories leaving the matching sphere at a direction given by
$(\vartheta_i,\varphi_i)$ and returning to it at $(\vartheta_f,\varphi_f)$. The
condition that $\underline G(r_0,r_0)$ obeys incoming-wave boundary conditions at
the final radius translates into the condition that only orbits going out
from the matching sphere into the long-range region and then returning to
$r_0$ are to be included, whereas orbits passing through the core region
are omitted. If all factors in the integrand except for the exponential are
assumed to vary slowly, the stationary-phase approximation reads
\begin{equation}
  \label{SclGreenSP}
  \begin{split}
  &G^{\rm scl}_{kk'} (r_0,r_0;E) = 2\pi r_0^2
    \sum_{\rm i\to f} \sin\vartheta_i\sin\vartheta_f \, \\
     & \qquad\times
        Y_k^\ast(\vartheta_f,\varphi_f) Y_{k'}(\vartheta_i,\varphi_i)\;
	\frac{\sqrt{|D_{\rm s.p.}|}}
          {\sqrt{\left|\displaystyle\det
              \frac{\partial^2 S}
                   {\partial(\vartheta',\varphi',\vartheta,\varphi)^2}
           \right|}} \\
     &\qquad\times\exp
        \left({\ri S(r_0,r_0) - \ri\frac{\pi}{2}(\sigma+\kappa)}\right) \;,
  \end{split}
\end{equation}
where $\kappa$ is the number of negative eigenvalues of the Hessian matrix
of $S$ occurring in the prefactor.

Because the initial state is assumed to be well localized, it is clear
that the outgoing waves generated by the photo-excitation originate in the
immediate neighborhood of the nucleus. Therefore, only trajectories
leaving the matching sphere radially need to be included in
(\ref{SclGreenSP}). By the same token, the trajectories can be assumed to
return to the matching radius radially. Thus, they are parts of closed
orbits starting precisely at the nucleus and returning there.

By transforming (\ref{GreenDDef}) to spherical coordinates and making use
of the relations
\begin{equation}
  \frac{\partial S}{\partial\vec x}=\vec p \;,\qquad
  \frac{\partial S}{\partial E}=t \;,
\end{equation}
the amplitude factor $D$ for radial trajectories can be simplified to
\begin{equation}
  D=-\frac{1}{\dot r \dot r'\,r^2 r'^2\sin\vartheta\sin\vartheta'}\,
     \det\frac{\partial(p_\vartheta',p_\varphi')}
              {\partial(\vartheta,\varphi)} \;.
\end{equation}
The determinants occurring in (\ref{SclGreenSP}) combine to
\begin{equation}
  \begin{split}
   &\det\frac{\partial(p_\vartheta',p_\varphi')}
             {\partial(\vartheta,\varphi)} \cdot
    \left(\det\frac{\partial^2 S}
                   {\partial(\vartheta',\varphi',\vartheta,\varphi)^2}
    \right)^{-1} \\[.5ex]
  =&\det\frac{\partial(p_\vartheta',p_\varphi',p_\vartheta,p_\varphi)}
             {\partial(\vartheta,\varphi,p_\vartheta,p_\varphi)} \cdot
    \left(\det\frac{\partial(-p_\vartheta',-p_\varphi',p_\vartheta,p_\varphi)}
                   {\partial(\vartheta',\varphi',\vartheta,\varphi)}
    \right)^{-1} \\[.5ex]
  =&\det\frac{\partial(\vartheta',\varphi')}
             {\partial(p_\vartheta,p_\varphi)} \;.
  \end{split}
\end{equation}
With these results, the Green's function matrix assumes the form
\begin{equation}
  \label{SclGreenChannel2}
  \begin{split}
  &G^{\rm scl}_{kk'} (r_0,r_0;E) = 2\pi \sum_{\rm c.o.}
    \frac{\sqrt{\sin\vartheta_i\sin\vartheta_f}}{\sqrt{|\dot r\dot r'|}}\\
  & \qquad\times
    \frac{Y_k^\ast(\vartheta_f,\varphi_f) Y_{k'}(\vartheta_i,\varphi_i)}
         {\sqrt{\left|\displaystyle
                      \det\frac{\partial(p_{\vartheta_f},p_{\varphi_f})}
                               {\partial(\vartheta_i,\varphi_i)}
                \right|}}\;
    \re^{\ri S(r_0,r_0) - \ri\pi(\sigma+\kappa)/2} \;.
  \end{split}
\end{equation}

The determinant in the denominator of (\ref{SclGreenChannel2}) measures the
dependence of the final angular momenta of the trajectory upon the starting
angles. As it stands, it suffers from the singularities of the spherical
coordinate chart: At the poles, neither the angle $\varphi$ nor the angular
momenta $p_\vartheta$ and $p_\varphi$ are well defined, so that close to
the poles, the calculation of the determinant becomes numerically
unstable. The determinant can be rewritten
in the form \cite{Bartsch03c}
\begin{equation}
  \label{MPrime}
  \det\frac{\partial(p_{\vartheta_f},p_{\varphi_f})}
           {\partial(\vartheta_i,\varphi_i)}
 =\sin\vartheta_i\sin\vartheta_f\,M
\end{equation}
with a $2\times 2$-determinant $M$ devoid of any singularities.  The
parameter $M$ was already used in \cite{Bartsch03a} to study the
bifurcations of closed orbits. We showed there that a closed orbit
bifurcates if and only if $M=0$.  With the form~(\ref{MPrime}) of the
stability determinant, the semiclassical Green's function matrix reads
\begin{equation}
  \label{SclGreenChannel3}
  \begin{split}
  G_{kk'}=2\pi\sum_{\rm c.o.}&\frac{1}{\sqrt{|\dot r\dot r'|}}\,
    \frac{Y_k^\ast(\vartheta_f,\varphi_f)Y_{k'}(\vartheta_i,\varphi_i)}
         {\sqrt{|M|}} \\
    &\times
    \exp\left(\ri S(r_0,r_0)-\ri\frac{\pi}{2}(\sigma+\kappa)\right) \;,
  \end{split}
\end{equation}
which is free of any singularities introduced by the spherical coordinates.

By virtue of (\ref{SGreen}), the semiclassical long-range scattering matrix
reads
\begin{equation}
  \label{SclSCrossed}
  \begin{split}
  &S_{kk'}^{\rm LR} = 2\ri\sum_{\rm c.o.} \frac{1}{\sqrt{|\dot r\dot r'|}}
    \frac{1}{f_k^{-}(r_0)}\frac{1}{f_{k'}^{-}(r_0)} \\
  &\qquad\times
    \frac{Y_k^\ast(\vartheta_f,\varphi_f)Y_{k'}(\vartheta_i,\varphi_i)}
         {\sqrt{|M|}} \;
    \re^{\ri S(r_0,r_0)-\ri\pi(\sigma+\kappa)/2} \;.
  \end{split}
\end{equation}
This expression can be further simplified if, for excited states close to
the ionization threshold, the radial wave functions
\begin{equation}
  \label{FradE0}
  f_l^{-}(r) \approx -\ri\sqrt{r} H^{(2)}_{2l+1}(\sqrt{8r})
\end{equation}
are approximated by the zero-energy wave functions, and the Hankel functions
are replaced with their asymptotic forms for large
arguments~\cite{Abramowitz}
\begin{equation}
  \label{HankelAsympt}
  H^{(2)}_{\nu}(x) \approx \sqrt{\frac{2}{\pi x}}
     \exp\left(-\ri x+\ri\frac{\pi}{2}\nu+\ri\frac{\pi}4\right) \;.
\end{equation}
This approximation has proven accurate in many cases of interest, but it
was called into question by Granger and Greene \cite{Granger00}. It will be
discussed further in section \ref{sec:CORot}, where we will show that there
is no reason to doubt its reliability.  It leads to
\begin{equation}
 \label{SCrossed0}
 \begin{split}
 &S_{lm,l'm'}^{\rm LR} = -2\pi\sum_{\rm c.o.} (-1)^{l+l'}
    \frac{Y_{lm}^\ast(\vartheta_f,\varphi_f)Y_{l'm'}(\vartheta_i,\varphi_i)}
         {\sqrt{|M|}} \\
    &\qquad\times
    \exp\left(\ri \left(S(r_0,r_0)+2\sqrt{8r_0}\right)
	-\ri\frac{\pi}{2}(\sigma+\kappa)\right) \;,
 \end{split}
\end{equation}
because, due to the conservation of energy, $\dot r^2/2=1/r$ if $E=0$. In
equation~\ref{SCrossed0}, the channel indices $k=(l,m)$ are finally written
out explicitly.

For a radial trajectory in a hydrogen atom going out from the nucleus to
$r=r_0$ at zero energy, the action is $\sqrt{8r_0}$, so that
\begin{equation}
  S_{\rm c.o.}=S(r_0,r_0)+2\sqrt{8r_0}
\end{equation}
is the action of a closed orbit, measured from its start at the nucleus to
its return. The semiclassical long-range $S$-matrix finally reads
\begin{equation}
  \label{SCrossed}
  \begin{split}
  S_{lm,l'm'}^{\rm LR} = -2\pi\sum_{\rm c.o.} &(-1)^{l+l'}
    \frac{Y_{lm}^\ast(\vartheta_f,\varphi_f)Y_{l'm'}(\vartheta_i,\varphi_i)}
         {\sqrt{|M|}} \\
    &\times
    \exp\left(\ri S_{\rm c.o.}-\ri\frac{\pi}{2}(\sigma+\kappa)\right) \;.
  \end{split}
\end{equation}
Both the action $S_{\rm c.o.}$ and the stability determinant $M$ are evaluated
at the nucleus rather than on the matching sphere. The response function is
given by
\begin{equation}
  \label{gCrossed}
  \begin{split}
  g^{\rm osc}(E) = 4\pi\sum_{\rm c.o.}
   &\frac{{\cal Y}^\ast(\vartheta_f,\varphi_f) {\cal Y}(\vartheta_i,\varphi_i)}
         {\sqrt{|M|}} \\
    &\times\exp\left(\ri S_{\rm c.o.}-\ri\frac{\pi}{2}\mu\right) \;,
  \end{split}
\end{equation}
where the Maslov index $\mu=\sigma+\kappa+1$ was increased by 1 to absorb an
additional phase, and the function
\begin{equation}
  \label{YCrossDef}
  {\cal Y}(\vartheta,\varphi)=\sum_{lm} (-1)^l d_{lm}
      Y_{lm}(\vartheta,\varphi) \;,
\end{equation}
with the core-region matrix elements $d_{lm}$ given by~(\ref{channelMat}),
characterizes the initial state and the exciting photon. Through the
$d_{lm}$, the function ${\cal Y}(\vartheta,\varphi)$ is
energy-dependent. In accordance with the choice of zero-energy radial wave
functions in the $S$-matrix elements, ${\cal Y}(\vartheta,\varphi)$ will be
evaluated at zero energy. This approximation has proven accurate in all
applications of closed-orbit theory considered in the literature so
far. However, from the $S$-matrix theory derivation it is obvious that the
energy-dependence of both the dipole matrix elements $d_{lm}$ and the
$S$-matrix elements can easily be included should the need arise.  The
semiclassical response function (\ref{gCrossed}) has the anticipated form
(\ref{resOscGen}) with
\begin{equation}
  \label{ACrossed}
  {\cal A}_{\rm c.o.} = 4\pi\;
   \frac{{\cal Y}^\ast(\vartheta_f,\varphi_f)\,{\cal Y}(\vartheta_i,\varphi_i)}
        {\sqrt{|M|}}
    \,\re^{\ri(\pi/2) \,\mu} \;.
\end{equation}

\subsection{Closed-orbit theory for symmetric systems}
\label{sec:CORot}

An atom in a single (electric or magnetic) external field possesses a
rotational symmetry around the field axis, which must be taken into account
in the derivation of the closed-orbit formulae.
The symmetry gives rise to a conserved magnetic quantum number
$m$, so that the angular momentum quantum number $l$ remains the only
relevant channel index. The
semiclassical scattering matrix reads \cite{Granger00}
\begin{equation}
  \label{Srot}
  \begin{split}
  &S_{ll'}^{\rm LR}=2^{3/2}\pi^{1/2}\sum_{\rm i\to f}
    \frac{\sqrt{|A|\sin\vartheta_i\sin\vartheta_f}}
         {|\dot r|f^-_l(r_0)f^-_{l'}(r_0)}\, \\
  &\qquad\times
    Y_{lm}^\ast(\vartheta_f,0) Y_{l'm'}(\vartheta_i,0) \;
    \re^{\ri S(r_0,r_0)-\ri\pi\tilde\mu/2-3\ri\pi/4} \;,
  \end{split}
\end{equation}
where
\begin{equation}
  \label{AGG}
  A = \left.\frac{\partial\vartheta_i}{\partial p_{\vartheta_f}}
      \right|_{p_{\vartheta_i}} \;,
\end{equation}
$\tilde\mu$ is the number of poles of $A$ encountered along the trajectory,
and the sum includes all classical trajectories with azimuthal angular
momentum $m$ joining the circles given
by polar angles $\vartheta_i$ and $\vartheta_f$ on the matching sphere. If the
radius of the matching sphere is much larger than the extent of the initial
state, the trajectories can again be assumed to leave the sphere and return
to it radially.
Strictly speaking, this condition can only be met if $m=0$, which we will
assume in what follows. If $m\ne 0$, the initial angular velocity
$\dot\varphi$ must be non-zero, but it will be small if the matching radius
is large. In this case, the trajectory will not actually close at the
nucleus, but swing by at a short distance.

Using, as above, the radial wave functions at zero energy, we obtain the
semiclassical scattering matrix
\begin{equation}
  \label{Srot2}
  \begin{split}
   &S_{ll'}^{\rm LR}=-(2\pi)^{3/2}(-1)^{l+l'}\ri \sum_{\rm c.o.}
    \sqrt{|A|\sin\vartheta_i\sin\vartheta_f}\, \\
   &\qquad\times
    Y_{lm}^\ast(\vartheta_f,0) Y_{l'm'}(\vartheta_i,0) \;
    \re^{\ri S_{\rm c.o.}-\ri\pi\tilde\mu/2-3\ri\pi/4}
  \end{split}
\end{equation}
and the response function
\begin{equation}
  \label{resRot}
  \begin{split}
  g^{\rm osc}(E) = 2(2\pi)^{3/2} \sum_{\rm c.o.}
   &\sqrt{|A|\sin\vartheta_i\sin\vartheta_f}\,
    {\cal Y}^\ast(\vartheta_f) {\cal Y}(\vartheta_i) \\
   &\times
    \exp\left(\ri S_{\rm c.o.}-\ri\frac{\pi}2\mu+\ri\frac{\pi}4\right)
  \end{split}
\end{equation}
with $\mu=\tilde\mu+2$ and
\begin{equation}
  \label{YRotDef}
  {\cal Y}(\vartheta)=\sum_l (-1)^l d_l Y_{lm}(\vartheta,0) \;.
\end{equation}
This result has the form (\ref{resOscGen}) with
\begin{equation}
  \label{ARot}
  \begin{split}
  {\cal A}_{\rm c.o.} = 2(2\pi)^{3/2}
   &\sqrt{|A|\sin\vartheta_i\sin\vartheta_f}\,
    {\cal Y}^\ast(\vartheta_f) {\cal Y}(\vartheta_i) \\
   &\times
    \exp\left(\ri\frac{\pi}2\mu+\ri\frac{\pi}4\right) \;.
  \end{split}
\end{equation}
It differs from the result obtained previously by Du and Delos
\cite{Du88} in that in their work the amplitude factor $A$ of
(\ref{AGG}) is replaced with
\begin{equation}
  \label{ADD}
  A_1 = \sqrt{\frac{2}{r_0}} 
    \left.\frac{\partial \vartheta_i}{\partial\vartheta_f}
    \right|_{p_{\vartheta_i}}
  \;.
\end{equation}
This discrepancy was noted and numerically investigated by Granger and
Greene \cite{Granger00}. They attribute it to the approximation of using
zero-energy wave functions, which can easily be avoided in the $S$-matrix
theory, but is an integral part of the derivation given by Du and Delos.

\begin{figure}
  \includegraphics[width=\columnwidth]{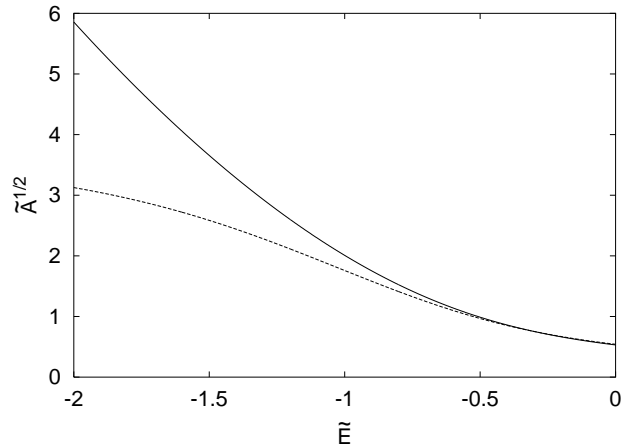}
  \caption{Scaled semiclassical amplitude factors after Granger and Greene
  \cite{Granger00}
  (Equation~(\ref{AGG}), solid line) and after Du and Delos \cite{Du88}
  (Equation~(\ref{ADD}), dashed line)
  for the closed orbit
  perpendicular to the magnetic field as a function of the scaled energy.
  The scaled matching radius is $\tilde r_0=0.01$.}
  \label{AmpEFig}
\end{figure}

For the closed orbit perpendicular to the field in the diamagnetic Kepler
problem and a scaled matching radius of $\tilde r_0=0.01$, the amplitudes
(\ref{AGG}) and (\ref{ADD}) are plotted in figure \ref{AmpEFig}. This
figure is similar to figure 1 in
\cite{Granger00}, although for the latter the matching radius is not
given. The agreement is excellent at scaled energies close to zero, but
becomes poor if the energy decreases. However, contrary to their
conclusions, the lack of agreement is not due to the zero-energy
approximation, but rather to the dependence of the amplitudes on the
matching radius.

This statement can be verified most conveniently if the motion is described
in semiparabolic coordinates
\begin{equation}
  \label{SParCoord}
  \mu=\sqrt{r+z}\;, \qquad \nu=\sqrt{r-z} \;.
\end{equation}
If the trajectory is recorded as a function of a parameter $\tau$ related
to the time $t$ by
\begin{equation}
  dt = 2r\,d\tau \;,
\end{equation}
and a prime denotes differentiation with respect to $\tau$,
for trajectories with vanishing azimuthal angular momentum
the equations of motion in the Coulomb region read
\begin{equation}
  \label{SPEqns}
  \begin{alignedat}{2}
    \mu' &= p_\mu \;, & \qquad \nu' &= p_\nu \;, \\
    p_\mu' &= 2E\mu \;, & \qquad p_\nu' &= 2E\nu \;.
  \end{alignedat}
\end{equation}
These equations are devoid of any singularities, so that they can
conveniently be used to discuss the motion close to the nucleus.
The transformation inverse to (\ref{SParCoord}) is given by
\begin{equation}
  \label{SPInverse}
    r = \frac 12\left(\mu^2+\nu^2\right) \;,  \qquad 
    \vartheta = \arccos\frac{\mu^2-\nu^2}{\mu^2+\nu^2}\;.
\end{equation}
The momenta transform according to
\begin{equation}
  \label{SPMomenta}
  p_r=\frac{\mu p_\mu+\nu p_\nu}{\mu^2+\nu^2} \;, \qquad
  p_\vartheta=\frac{\mu p_\nu-\nu p_\mu}{2\operatorname{sign}(\mu\nu)}\;.
\end{equation}
Note that the transformation from semiparabolic to Cartesian coordinates is
not one-to-one, but that $\mu$ and $\nu$ are fixed up to the choice of sign
only.

To evaluate (\ref{AGG}) and (\ref{ADD}), the derivatives $\partial
p_{\vartheta_f}/\partial\vartheta_i$ and
$\partial\vartheta_f/\partial\vartheta_i$, must be calculated and their
dependence on the matching radius $r$ must be determined. As the radial
trajectory specified by a starting angle $\vartheta_i$ is independent of
the radius where the angle is measured, the $r$-dependence of the
derivatives is determined by the returning trajectories only. It can be
evaluated as follows:

We arbitrarily fix the returning time of a closed orbit at $\tau=0$, so that
$\mu(0)=\nu(0)=0$. The solution to (\ref{SPEqns}) describing a trajectory
returning at an angle $\vartheta_f$ is given by
\begin{equation}
  \label{COReturn}
  \begin{aligned}
    \mu(\tau) &= 2\,\frac{\cos(\vartheta_f/2)}{\sqrt{-2E}}\,
                    \sin\left(\sqrt{-2E}\,\tau\right)
              = -\sqrt{2r}\cos\frac{\vartheta_f}2 \;, \\
    \nu(\tau) &= 2\,\frac{\sin(\vartheta_f/2)}{\sqrt{-2E}}\,
                    \sin\left(\sqrt{-2E}\,\tau\right)
              = -\sqrt{2r}\sin\frac{\vartheta_f}2 \;, \\
    p_\mu(\tau) &=2\cos\frac{\vartheta_f}2
                    \cos\left(\sqrt{-2E}\,\tau\right)
                =2\sqrt{1+Er}\cos\frac{\vartheta_f}2\;, \\
    p_\nu(\tau) &=2\sin\frac{\vartheta_f}2
                    \cos\left(\sqrt{-2E}\,\tau\right)
                =2\sqrt{1+Er}\cos\frac{\vartheta_f}2\;,
  \end{aligned}
\end{equation}
where the coefficients were chosen to satisfy the conservation of energy
and to give the correct returning angle after a transformation to Cartesian
coordinates. The second expression in each line follows from
$\mu^2+\nu^2=2r$, whence for $\tau<0$
\begin{equation}
\begin{gathered}
  \label{scr}
  \sin\left(\sqrt{-2E}\,\tau\right)=-\sqrt{-Er} \;, \\
  \cos\left(\sqrt{-2E}\,\tau\right)=\sqrt{1+Er} \;.
\end{gathered}
\end{equation}

Equations of motion for the derivatives $\partial\mu/\partial\vartheta_i$ and
$\partial\nu/\partial\vartheta_i$ are obtained by linearizing
(\ref{SPEqns}). Since (\ref{SPEqns}) is already linear, the derivatives
satisfy the same equations of motion as the coordinates themselves as long
as the electron moves in the Coulomb region. There the solutions read
\begin{equation}
  \frac{\partial\mu}{\partial\vartheta_i} = 
    \frac{a_\mu}{\sqrt{-2E}}\sin\left(\sqrt{-2E}\,\tau\right)
    +b_\mu\cos\left(\sqrt{-2E}\,\tau\right)
\end{equation}
and
\begin{equation}
  \begin{split}
  \frac{\partial p_\mu}{\partial\vartheta_i} &=
    \frac{d}{d\tau}\,\frac{\partial\mu}{\partial\vartheta_i}\\ &=
    a_\mu\cos\left(\sqrt{-2E}\,\tau\right)
    -\sqrt{-2E}\,b_\mu\sin\left(\sqrt{-2E}\,\tau\right) \;.
  \end{split}
\end{equation}
Equation (\ref{scr}) yields
\begin{equation}
  \label{muVar}
  \begin{aligned}
    \frac{\partial\mu}{\partial\vartheta_i} &= 
      -a_\mu\sqrt{\frac r2} + b_\mu\sqrt{1+Er} \;, \\
    \frac{\partial p_\mu}{\partial\vartheta_i} &=
      a_\mu\sqrt{1+Er}-\sqrt{2r}Eb_\mu \;,
  \end{aligned}
\end{equation}
so that the coefficients
\begin{equation}
  a_\mu=\frac{\partial p_{\mu_f}}{\partial\vartheta_i} \;,\qquad
  b_\mu=\frac{\partial\mu_f}{\partial\vartheta_i}
\end{equation}
can be identified with the values of the derivatives obtained at
$r=0$. Analogous expressions hold for $\partial\nu/\partial\vartheta_i$.

From (\ref{SPMomenta}), the amplitude (\ref{AGG})
\begin{equation}
  \begin{split}
  \frac{1}{A} &= \frac{\partial p_\vartheta}{\partial\vartheta_i} \\
    &=
    \frac{1}{2\operatorname{sign}(\mu\nu)}
    \left(\mu\frac{\partial p_\nu}{\partial\vartheta_i} +
          p_\nu\frac{\partial\mu}{\partial\vartheta_i} -
          \nu\frac{\partial p_\mu}{\partial\vartheta_i} -
          p_\mu\frac{\partial\nu}{\partial\vartheta_i}
    \right) \\
    &= \frac{1}{2\operatorname{sign}(\mu\nu)}
     \left(\frac{\partial\mu_f}{\partial\vartheta_i}\,p_{\nu_f} - 
           \frac{\partial\nu_f}{\partial\vartheta_i}\,p_{\mu_f}
     \right)
  \end{split}
\end{equation}
can be evaluated. It is independent of $r$, as could have been anticipated
from the fact that $p_\vartheta$ is a component of the total angular
momentum and thus is conserved along the trajectory once the electron has
entered the Coulomb region. The amplitude $A^{-1}$ can also, up to an
immaterial choice of sign, be identified with the monodromy matrix element
\begin{equation}
  \label{m12Def}
  m_{12}=\frac12\left(
           \frac{\partial\nu_f}{\partial\vartheta_i}\,p_{\mu_f}-
           \frac{\partial\mu_f}{\partial\vartheta_i}\,p_{\nu_f}
         \right)
\end{equation}
introduced by Bogomolny~\cite{Bogomolny89} to describe the semiclassical
amplitudes, so that the amplitudes derived by Granger and Greene from the
$S$-matrix theory agree with Bogomolny's.

\begin{figure}
  \includegraphics[width=\columnwidth]{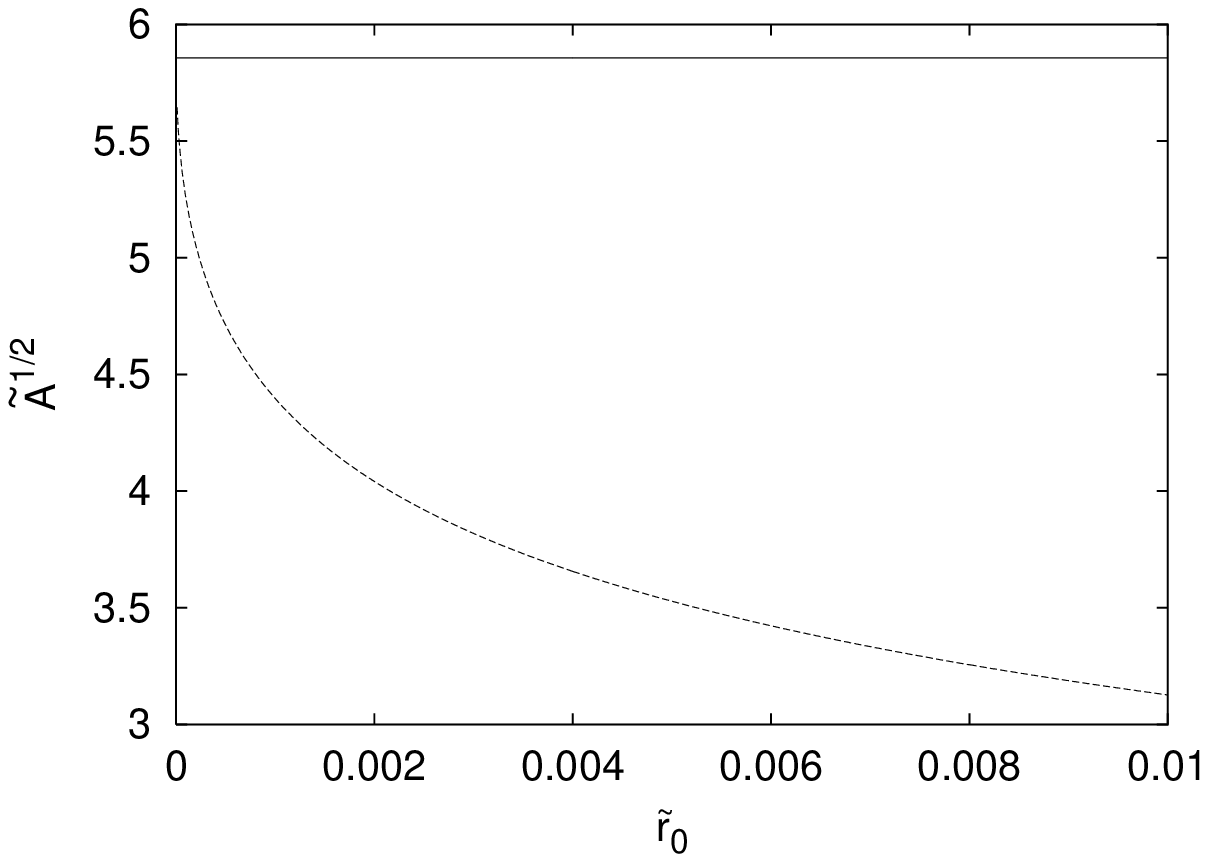}
  \caption{Scaled semiclassical amplitude factors after Granger and Greene
  \cite{Granger00}
  (Equation~(\ref{AGG}), solid line) and after Du and Delos \cite{Du88}
  (Equatiion~(\ref{ADD}), dashed line)
  for the closed orbit perpendicular to the magnetic field as a function of
  the matching radius at $\tilde E=-2$.}
  \label{AmpRFig}
\end{figure}

Similarly, the amplitude (\ref{ADD}) used by Du and Delos reads, by
(\ref{SPInverse}),
\begin{equation}
  \begin{split}
    \frac{1}{A_1} &= \sqrt{\frac{r}{2}}\,
                     \frac{\partial\vartheta}{\partial\vartheta_i} \\
    &= \frac{\operatorname{sign}(\mu\nu)}{2} \left[
         p_{\mu_f}\left(
            \sqrt{\frac{r}{2}}\,\frac{\partial p_{\nu_f}}{\partial\vartheta_i}-
            \sqrt{1+Er}\,\frac{\partial\nu_f}{\partial\vartheta_i}
          \right)\right. \\
       & \left.\hspace{2cm}-p_{\nu_f}\left(
            \sqrt{\frac{r}{2}}\,\frac{\partial p_{\mu_f}}{\partial\vartheta_i}-
            \sqrt{1+Er}\,\frac{\partial\mu_f}{\partial\vartheta_i}
          \right)
       \right] \\
    &= \frac{1}{A} + {\cal O}\left(\sqrt{r}\right) \;.
  \end{split}
\end{equation}
Thus, the amplitudes $A$ and $A_1$ agree in the limit of vanishing matching
radius, but the amplitude $A_1$ proposed by Du and Delos exhibits a strong
dependence on $r$, whereas the amplitude $A$ given by Granger and Greene
does not. These findings can also be confirmed numerically. Figure
\ref{AmpRFig} shows the two amplitudes for the closed orbit perpendicular
to the magnetic field at a scaled energy of $\tilde E=-2$ as a function of
the scaled matching radius $\tilde r_0$.  The dependence of $A_1$ on
$\tilde r_0$ is considerable.

We have thus shown that, contrary to the conclusion reached by Granger and
Greene,  the discrepancy between their semiclassical amplitude and that
obtained by Du and Delos is not due to the zero-energy approximation, but
rather due to the choice of a finite matching radius. In addition, the
amplitude derived by Granger and Greene is not specific to the $S$-matrix
formulation, it agrees with the result derived earlier by Bogomolny in the
context of a semiclassical wave function formalism. Nevertheless, as it
eliminates the need to specify a finite matching radius and
allows one to calculate all classical quantities at the nucleus, it
 seems more appropriate than the amplitude given by Du and Delos,
which introduces a certain arbitrariness in the choice of a matching
radius.

\section{The scaled quantum spectrum}
\label{sec:Xqm}

If Schr\"odinger's equation for the crossed-fields hydrogen atom is
rewritten in terms of the scaled energy and the scaled electric field
strength, a quadratic eigenvalue problem for the scaling parameter $w$ is
obtained. An exact numerical method of solution for the quadratic
eigenvalue problem has become available only recently \cite{Rao01}.  We
resort to the method introduced by Main \cite{Main99d}, which relies on an
approximate linearization of the eigenvalue problem to compute eigenvalues
in a small spectral interval. The accuracy of the linearization can be
verified by comparing results calculated using different overlapping
intervals. The eigen\-values are obtained to a relative accuracy of at
least $10^{-7}$, which is far beyond the typical accuracy of semiclassical
approximations, so that the algorithm is well suited to this study.

\begin{figure}
  \includegraphics[width=\columnwidth]{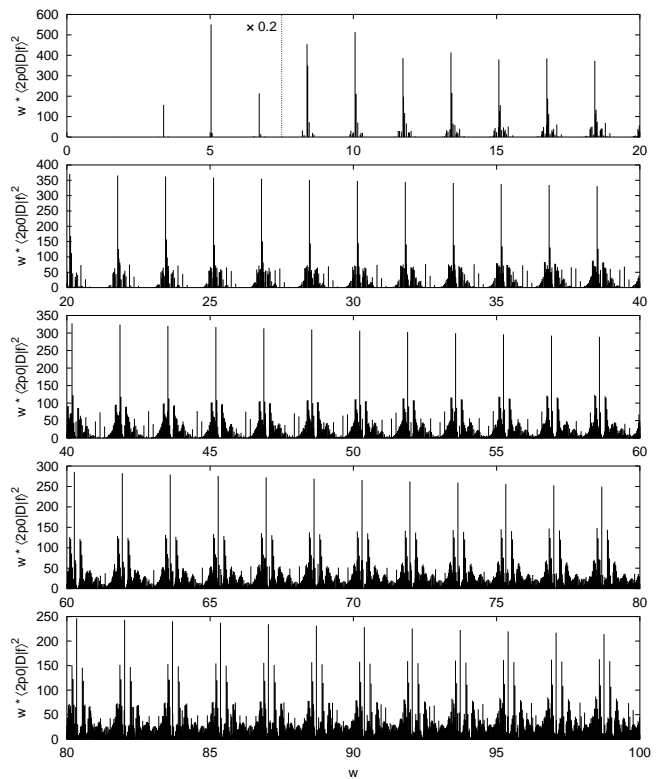} 
  \caption{Quantum photo-absorption spectrum at the scaled energy $\tilde
  E=-1.4$ and the scaled electric field strength $\tilde F=0.1$. The
  initial state is $|2p0\rangle$, the light is polarized along the magnetic
  field axis. The plot shows the squared dipole matrix elements, which
  for graphical reasons are multiplied by $w$. The strengths of the
  extraordinarily strong lines of the lowest $n$-manifolds at $w<7.5$ 
  are scaled down by a factor of 0.2.}
  \label{QuantumFig}
\end{figure}

In the following we will discuss quantum and semiclassical photo-absorption
spectra obtained for the scaled energy $\tilde E=-1.4$ and the scaled
electric field strength $\tilde F=0.1$ with the initial state $|2p0\rangle$
and light linearly polarized along the magnetic field axis. A quantum
spectrum for these parameter values is shown in figure~\ref{QuantumFig}. As
for a semiclassical analysis (see section~\ref{sec:SclRec}) it is essential
to have as many eigenvalues available as possible, the calculation was
extended up to $w=100$. The spectrum shown in figure~\ref{QuantumFig}
contains nearly 30,000 lines, many of which are too weak to be discernible
in the plot.

The eigenenergies of the field-free hydrogen atom satisfy
\begin{equation}
  E=w^{-2}\tilde E=-\frac{1}{2n^2} \;,
\end{equation}
so that in the scaled spectrum the unperturbed $n$-manifolds appear
equidistantly spaced at
\begin{equation}
  w=\sqrt{-2\tilde E}\,n \;.
\end{equation}
These spacings can clearly be recognized in figure~\ref{QuantumFig}. At low
values of $w$, neighboring $n$-manifolds are isolated. Furthermore, in
this region the magnetic quantum number $m$ is nearly conserved. This is
apparent from the fact that each $n$-manifold contains a central group of
strong levels corresponding to $m=0$, which can be excited even at $\tilde
F=0$, and adjacent groups of considerably weaker levels with $m=\pm
1$. Levels with higher magnetic quantum numbers are too weak in this region
to be seen in the figure. At higher values of $w$, the conservation of $m$
is violated, and individual $n$-manifolds acquire strong side bands.  At
even higher $w$, different $n$-manifolds strongly overlap. Throughout the
spectral range shown, groups of strong lines indicating the centers of
different $n$-manifolds are clearly discernible.

\section{Low-resolution semiclassical spectra}
\label{sec:SclLo}

A semiclassical approximation to a scaled photo-absorp\-tion spectrum is
obtained if the closed-orbit theory formulae of section~\ref{sec:COCrossed}
are rewritten in terms of scaled quantities, viz.
\begin{equation}
  \label{gOscScal}
  g^{\rm osc}(w) = \frac{1}{w}\sum_{\rm c.o.}{\cal \widetilde A}_{\rm c.o.}
    \exp\left(\ri w \tilde S_{\rm c.o.}\right)
\end{equation}
with
\begin{equation}
  {\cal\widetilde A}_{\rm c.o.} = 4\pi
    \frac{{\cal Y}^\ast(\vartheta_f,\varphi_f)\,
          {\cal Y}(\vartheta_i,\varphi_i)}
         {\sqrt{|\tilde M|}}
    \,\re^{\ri(\pi/2) \,\mu} \;.
\end{equation}

When low-resolution photo-absorption spectra are to be calculated from
(\ref{gOscScal}), a method of cut-off must be adopted to deal with the
divergence of the semiclassical closed-orbit sum. For this section, we choose
a Gaussian cut-off, i.e. (\ref{gOscScal}) is replaced with
\begin{equation}
  \label{gOscSmooth}
  g_\sigma^{\rm osc}(w) = 
    \frac{1}{w}\sum_{\rm c.o.}{\cal \widetilde A}_{\rm c.o.}
    \exp\left(\ri w \tilde S_{\rm c.o.}
             -\frac{\tilde S_{\rm c.o.}^ 2}{2\sigma^2}\right)\;,
\end{equation}
so that orbits with scaled actions larger than the cut-off action $\sigma$
are smoothly suppressed. This smoothing corresponds to a convolution of the
quantum signal with a Gaussian of width $1/\sigma$.

\begin{figure}
  \includegraphics[width=\columnwidth]{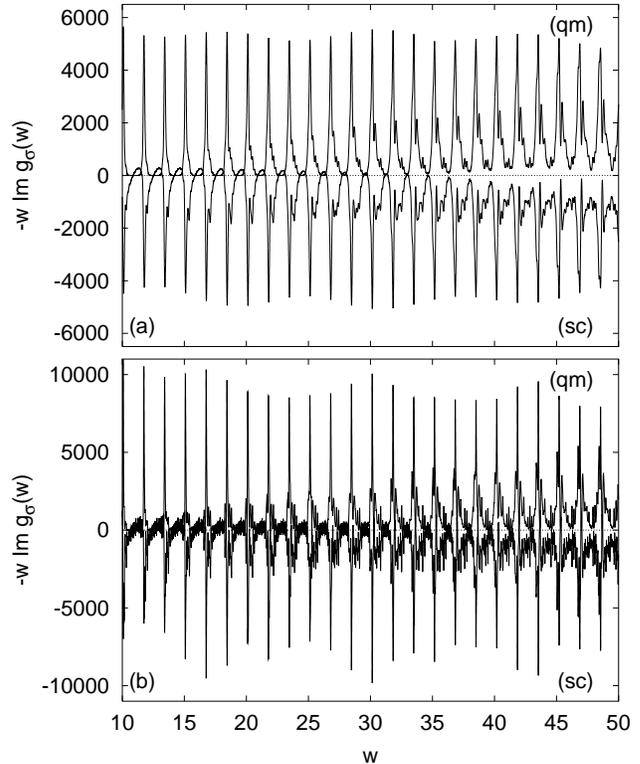}
  \caption{Smoothed quantum
  (upper halves) and semiclassical (lower halves, inverted) photo-absorption
  spectra with cut-off action (a) $\sigma=20$ and (b) $\sigma=50$.}
  \label{XLoFig}
\end{figure}

To facilitate the comparison of (\ref{gOscSmooth}) with the convoluted
quantum spectrum, we added the smooth part of the spectrum to
$g_\sigma^{\rm osc}$, which was calculated by convoluting the quantum
spectrum with a Gaussian of width $1/\sigma=1$. This function is broad
enough to wipe out the distinction between neighboring principal quantum
numbers. Results obtained for $\sigma=20$ and $\sigma=50$ are shown in
figure~\ref{XLoFig}. In both cases it is apparent that the large-scale
structure of equidistant principal quantum numbers is well reproduced by
the semiclassical approximation. In the quantum spectra, the substructure
of the individual $n$-shells can be discerned to a certain degree, given by
the smoothing width $1/\sigma$. In the case of $\sigma=20$, much of this
fine structure is also present in the semiclassical spectrum, but often the
agreement is not good quantitatively. In particular, the peaks
corresponding to the lowest $n$-manifolds are considerably wider in the
semiclassical than in the quantum spectrum.

If the cut-off action is increased to $\sigma=50$, finer details are
resolved in the quantum spectrum. At the same time, the semiclassical
closed-orbit sum becomes more oscillatory to reproduce this fine
structure. It appears, however, to be somewhat over-oscillatory, developing
structures absent from the quantum spectrum. This type of behavior is
typical of closed-orbit sums in non-integrable systems. Thus, it can be
questioned if the low-resolution closed-orbit sum can meaningfully be
extended to even longer orbits. A high-resolution quantization based on the
present semiclassical approximation will be presented in the following
section.

\section{High-resolution semiclassical spectra}
\label{sec:SclHi}

For the calculation of a scaled semiclassical spectrum, the method of
semiclassical quantization by harmonic inversion of $\delta$ function
signals \cite{Main98b,Bartsch01} can be applied. This technique
requires the inclusion of closed orbits up to a maximum scaled action,
i.e. it replaces the Gaussian cut-off used for the low-resolution
semiclassical spectra presented in the previous section with a rectangular
cut-off. A rough estimate for the required cut-off action can be obtained
by means of perturbation theory \cite{Bartsch02}. 
\begin{equation}
  \label{SMaxPert}
  \tilde S_{\rm max} = -8\pi\tilde E\,n \;.
\end{equation}
For the case $\tilde
E=-1.4$ and $n=9$, i.e. $w=15.06$, this estimate yields $\tilde S_{\rm
max}/2\pi \approx 50$.

According to (\ref{SMaxPert}), to compute levels at high quantum numbers
$n$ a long semiclassical signal is needed, which can be hard or even
impossible to obtain. We calculated closed orbits up to $\tilde S_{\rm
max}/2\pi=200$, so that the orbital data is available for nearly 18,000
closed-orbit multiplets. However, for reasons to be described in
section~\ref{sec:SclRec} a useful semiclassical signal can be constructed
up to $\tilde S_{\rm max} /2\pi\approx 60-70$ only, so that, from the above
estimate, the semiclassical calculation cannot reach manifolds much higher
than $n=10$.  On the other hand, the semiclassical approximation must be
expected to yield more accurate results for higher quantum numbers. Thus,
when a high-resolution semiclassical spectrum is to be calculated, a
compromise must be made between the contradictory requirements of
describing a spectral region at sufficiently high quantum numbers and with
a sufficiently low spectral density.

\begingroup
\squeezetable
\begin{table*}
\begin{ruledtabular}
\begin{minipage}{.48\textwidth}
\begin{tabular}{@{}r|rrr@{}}
   $n$ &$w_f$ (scl.)& $w_f$ (qm.) & $\langle 2p0|D|f\rangle^2$  \\
\hline
     &           &     9.88321  &  1.3617 \\
\phantom{10}				    
     &           &     9.91431  &  3.1145 \\
     &           &     9.97747  &  1.7474 \\
     &  10.05366 &    10.05912  & 51.0512 \\
  6  &  10.09551 &    10.09621  & 20.9313 \\
     &  10.15461 &    10.15378  &  7.0060 \\
     &           &    10.24076  &  0.9608 \\
     &           &    10.26612  &  2.0777 \\
     &           &    10.31803  &  1.9385 \\
\hline				 	    
				 	    
     &           &    11.56497  &  2.5663 \\
     &  11.60898 &    11.60820  &  2.5875 \\
     &  11.66889 &    11.67341  &  2.3104 \\
     &  11.72048 &    11.73128  & 32.8808 \\
  7  &           &    11.75121  & 16.7278 \\
     &           &    11.78850  & 10.0092 \\
     &           &    11.84856  &  5.6249 \\
     &           &    11.92188  &  1.9229 \\
     &           &    11.95821  &  1.7923 \\
     &           &    12.01338  &  2.4821 \\
\hline				 	    
				 	    
     &           &    13.23441  &  1.3668 \\
     &           &    13.25629  &  2.5141 \\
     &           &    13.30255  &  1.9971 \\
     &  13.36921 &    13.36913  &  2.8189 \\
     &  13.40177 &    13.40568  & 30.8875 \\
 8   &  13.44313 &    13.43744  & 16.0829 \\
     &  13.48737 &    13.48146  &  4.8263 \\
     &           &    13.54340  &  4.3111 \\
     &           &    13.59258  &  1.0747 \\
     &           &    13.61133  &  1.9475 \\
     &           &    13.65111  &  1.4081 \\
     &           &    13.70866  &  2.9676 \\
\hline				 	    
				 	    
     &           &    14.91192  &  2.1880 \\
     &           &    14.94654  &  2.9922 \\
     &           &    14.99711  &  1.4563 \\
 9   &  \raisebox{-1.5ex}[2ex]{15.06960} &    15.06470  &  3.2226 \\
     &           &    15.07888  & 25.1866 \\
     &           &    15.10074  &  8.4317 \\
  \multicolumn{2}{c}{}
\end{tabular}
\end{minipage}
\hfill
\begin{minipage}{.48\textwidth}
\begin{tabular}{@{}r|rrr@{}} 		 
   $n$ &$w_f$ (scl.)& $w_f$ (qm.) & $\langle 2p0|D|f\rangle^2$ \\
\hline
     &  15.12905 &    15.12748  & 10.3140 \\
     &  15.17892 &    15.17491  &  2.2476 \\
     &  15.23623 &    15.23830  &  3.1064 \\
 9   &  15.26111 &    15.27005  &  1.7749 \\
     &           &    15.30024  &  2.3710 \\
     &           &    15.34449  &  1.0296 \\
     &           &    15.40389  &  3.3462 \\
\hline				 	    
					    
     &           &    16.57908  &  0.7173 \\
     &           &    16.58435  &  1.7007 \\
     &           &    16.60357  &  1.7437 \\
     &  16.64355 &    16.63843  &  2.9662 \\
     &  16.69069 &    16.69180  &  0.9974 \\
     &  16.74965 &    16.75258  & 22.9143 \\
     &           &    16.76016  &  3.4901 \\
 10  &  16.78346 &    16.78269  & 11.1809 \\
     &  16.81329 &    16.81827  &  6.6898 \\
     &           &    16.86870  &  0.9825 \\
     &  16.93431 &    16.93323  &  2.0584 \\
     &           &    16.94303  &  1.4143 \\
     &           &    16.96000  &  1.4406 \\
     &           &    16.99085  &  2.3893 \\
     &           &    17.09909  &  3.5870 \\
     &           &    17.25847  &  0.7647 \\
\hline				 	    
				 	    
     &           &    18.25950  &  2.1201 \\
     &           &    18.27572  &  0.9781 \\
     &           &    18.29004  &  2.6665 \\
     &           &    18.33096  &  2.7709 \\
     &  18.42131 &    18.42600  & 20.2420 \\
     &           &    18.45136  &  6.1451 \\
     &           &    18.45555  &  3.5970 \\
 11  &  18.47472 &    18.47149  &  7.2231 \\
     &           &    18.50996  &  4.0510 \\
     &           &    18.61835  &  1.7975 \\
     &           &    18.62818  &  1.2089 \\
     &           &    18.64563  &  2.2348 \\
     &           &    18.68226  &  2.2558 \\
     &           &    18.79427  &  3.6707 \\
     &  18.93585 &    18.95442  &  1.0263 \\
\end{tabular}
\end{minipage}
\end{ruledtabular}
\caption{Semiclassical and quantum eigenvalues $w_f$ of the scaling
parameter for $\tilde E=-1.4$ and $\tilde F=0.1$. See text for a detailed
description. The dipole matrix elements $\langle 2p0|D|f\rangle^2$ were
obtained from a quantum spectrum.}
\label{XSclTab}
\end{table*}
\endgroup

For the harmonic analysis of the closed-orbit sum we applied the method of
$\delta$ function decimated signal diagonalization
\cite{Main00a,Bartsch01}, which yields not only semiclassical eigenvalues
and amplitudes, but also an error parameter estimating the precision of the
eigenvalues.  Results obtained for $\tilde E=-1.4$ and $\tilde F=0.1$ with
a signal length of $\tilde S_{\rm max} /2\pi=60$ are compiled in
table~\ref{XSclTab}. The table contains the quantum eigen\-values of $w$
and their dipole matrix elements for levels satisfying $\langle
2p0|D|f\rangle^2>0.7$. It is obvious at a glance that out of the multitude
of spectral lines with intensities varying over many orders of magnitude
(most of which are not contained in the table) only the strongest lines
were detected in the semiclassical spectrum. The semiclassical eigenvalues
given are characterized by having small imaginary parts, small error
parameters and large amplitudes as well as being stable with respect to a
variation of numerical parameters. The calculation operates at the edge of
convergence, and in a few cases one can be in doubt whether a level should
be included according to these fairly ``soft'' criteria, but in general a
clear decision can be made. Semiclassical values for the transition
strengths are not given because they are not reasonably well converged and
depend strongly on the numerical parameters.

One might expect that in each $n$-manifold it is the strongest lines that
are detected semiclassically, and in general this expectation is confirmed
by the numerical data. This can clearly be seen, e.g., in the manifold
$n=6$, which contains the most stably converged lines in the
spectrum. There are, however, a few conspicuous exceptions, e.g. at $n=7$,
where strong lines are missing whereas comparatively weak lines are
found. For $n=5$, no lines at all can be
computed from the given semiclassical signal. If the signal length is
decreased to $\tilde S_{\rm max} /2\pi=50$, the three strongest lines
appear in the spectrum in this manifold.

At higher $n$, the number of strong lines in the quantum spectrum
increases. So does the number of lines in the semiclassical spectrum until
$n=11$, where only three semiclassical lines are found. They appear rather
arbitrarily scattered across the quantum spectrum, and their convergence is
notably worse than in lower manifolds. It is clear that in this $n$-shell
the semiclassical quantization with the given signal is about to break
down. At $n=12$, no lines can be detected semiclassically.  As, from the
above discussion, this failure was to be expected because the required
signal length becomes too large, the obvious way to improve convergence
seems to be to use a longer signal. However, if the signal length is
increased to $\tilde S_{\rm max} /2\pi=70$, no reasonably converged
semiclassical lines can be found in any $n$-manifold. Neither are results
improved if the technique of harmonic inversion of cross-correlated
closed-orbit sums \cite{Main99a,Main99c} is applied. This method has proven
powerful in reducing the signal length required in a semiclassical
quantization. In the present case, however, because the cross-correlation
increases the total number of frequencies obtained from the harmonic
inversion, the true eigenvalues are hidden among a multitude of spurious
frequencies, and no useful results can be obtained.

For the time being, therefore, the results given in table~\ref{XSclTab}
represent what can be achieved in the semiclassical quantization of the
crossed-fields hydrogen atom. They confirm the applicability of the
closed-orbit theory approach in principle, but they also reveal a
fundamental problem in its present formulation. From the analysis of the
ideal test signal it is clear that the signal length available is
sufficient for a stable signal analysis. Thus, if the semiclassical results
are not good, the semiclassical signal itself, rather than the signal
analysis, must be to blame. This conclusion is confirmed by the observation
that an increased signal length destroys the results rather than improves
them. We therefore start searching for a flaw in the construction of the
semiclassical photo-absorption spectrum.

A conspicuous problem lies in the fact that the set of closed orbits
available is incomplete. In no series of rotators or vibrators can
arbitrarily long orbits be calculated. In the case of vanishing electric
field there is a critical angle $\vartheta_{\rm c}$ which the starting
angles of both rotators and vibrators approach as the orbits get
longer. This convergence indicates that the orbits approach a separatrix
between two families of tori in phase space. If sufficiently long orbits
are studied, there are many closed orbits with very similar initial
conditions, so that the numerical search for closed orbits must eventually
fail.

\begin{figure}
  \includegraphics[width=\columnwidth]{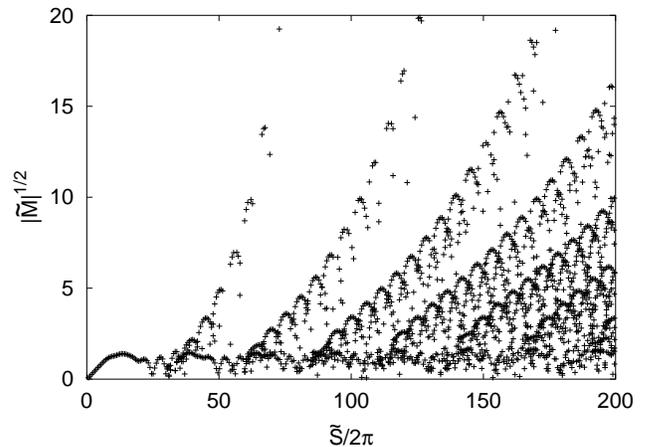}
  \caption{Stability determinants of vibrators as a function of the
  action for $\tilde E=-1.4$, $\tilde F=0.1$.}
  \label{StabMFig}
\end{figure}

The region of phase space where the unknown orbits are located is lying
close to a separatrix, so that it is highly unstable. The orbits can
therefore be expected not to contribute much to the semiclassical
signal. The magnitude of an orbit's contribution to the closed-orbit sum
(\ref{gCrossed}) is determined mainly by its stability determinant
$M$. Figure~\ref{StabMFig} shows the stability determinants of the vibrator
orbits for $\tilde E=-1.4$, $\tilde F=0.1$ as a function of the scaled
action. Different series of vibrators can clearly be discerned in the plot.
It is indeed unstable orbits with large $\tilde M$ that are missing in the
data set, but on the other hand the stability determinants of the missing
orbits are not large enough to regard the corresponding semiclassical
amplitudes as negligibly small. Because a vast majority of orbits has
small $\tilde M$ and was found, one can still hope that useful results
can be obtained from the semiclassical signal, at least for quantum states
not located in the separatrix region in phase space, but it is clear that the
quality of the semiclassical signal is reduced by its incompleteness.

To assess in detail the detrimental effect of the missing orbits and of any
other sources of error that may exist, we carry out a semiclassical analysis
of the quantum spectrum.

\section{Semiclassical recurrence spectra}
\label{sec:SclRec}

According to equation~(\ref{gOscScal}), in a scaled photo-absorp\-tion
spectrum every closed orbit contributes a purely sinusoidal modulation to
$w\,g(w)$. This contribution can be extracted from the spectrum either by a
conventional Fourier transform or by means of a high-resolution method. The
spectral analysis yields information about classical orbits returning to
the nucleus. For this reason, the transformed spectrum is referred to as a
recurrence spectrum.  High-resolution methods \cite{Main99d} extract the
scaled actions and scaled semiclassical amplitudes of individual orbits and
thus yield more complete information about the semiclassical spectrum than
the Fourier transform, but they fail if the average density of closed
orbits per unit of scaled action is too large. By contrast, due to its
linearity the Fourier transform can be applied to any part of the
recurrence spectrum with equal ease and numerical stability, irrespective
of the spectral density. In dense regions, it will not be able to identify
individual closed orbits, but it will nevertheless yield a recurrence
spectrum that can be compared to the classical data. In this section we
will present results obtained by both the Fourier transform and a
high-resolution method. The semiclassical recurrence spectra will be
compared to classical results in order to identify the reason why the
semiclassical signal is only partially suitable to a semiclassical
quantization.

Using either method, it is essential to note that the semiclassical
closed-orbit formula cannot be expected to yield accurate results for the
lowest levels. Thus, the low $n$-manifolds must be excluded from the
semiclassical analysis, i.e. the analysis is based on the quantum spectrum
given in an interval $[w_{\rm min},w_{\rm max}]$ instead of $[0,w_{\rm
max}]$.
Furthermore, to minimize the impact of boundary effects due to the finite
length of the semiclassical spectrum, a smooth Gaussian cut-off with width
$\kappa$ centered at $w_0=(w_{\rm min}+w_{\rm max})/2$ is
introduced. The smoothing replaces
the peaks of the semiclassical recurrence spectrum by
Gaussians of width $1/\kappa$. The recurrence spectra presented here were
calculated from the quantum spectrum shown in figure~\ref{QuantumFig}, for
$\tilde E=-1.4$ and $\tilde F=0.1$, with $w_{\rm min}=20$, $w_{\rm
max}=100$, and $\kappa=10$. For the high-resolution recurrence spectra, the
method of $\delta$ function decimated signal diagonalization was used.

\begin{figure}
  \includegraphics[width=.92\columnwidth]{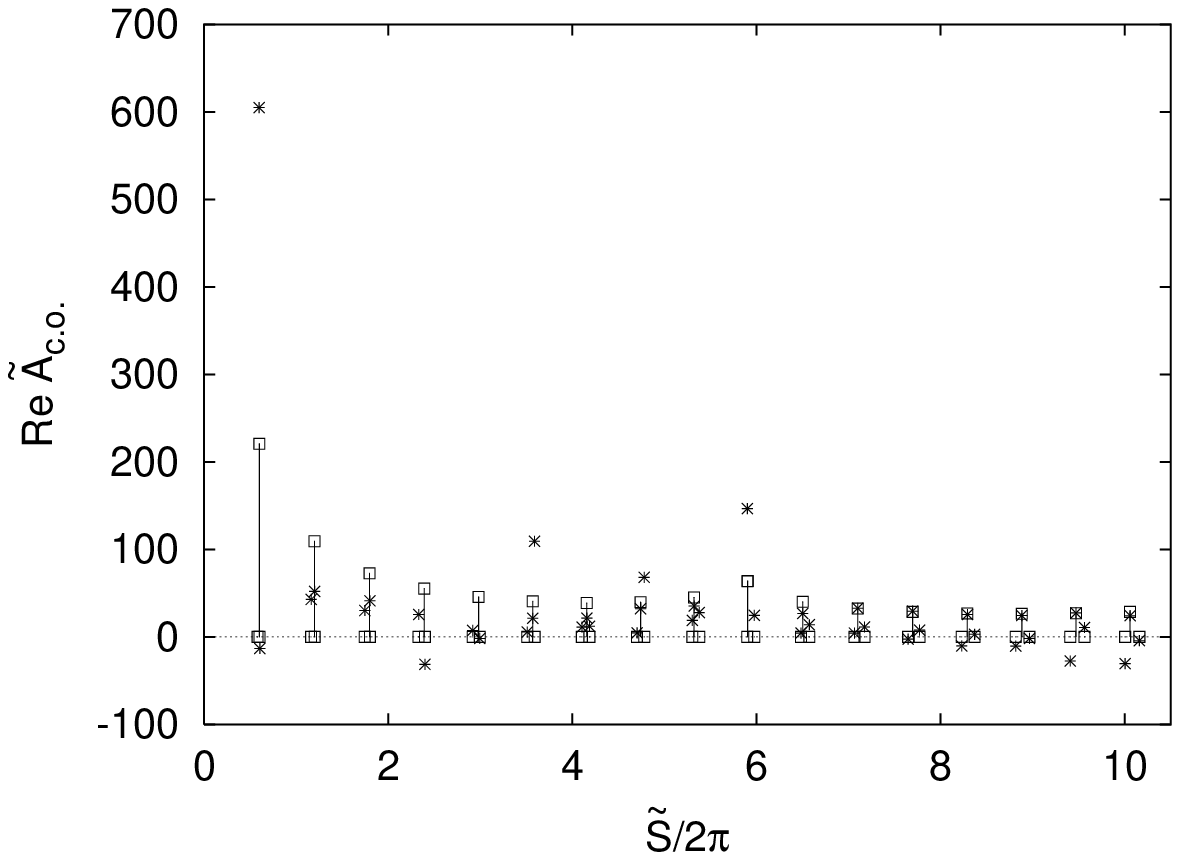}
  \includegraphics[width=.92\columnwidth]{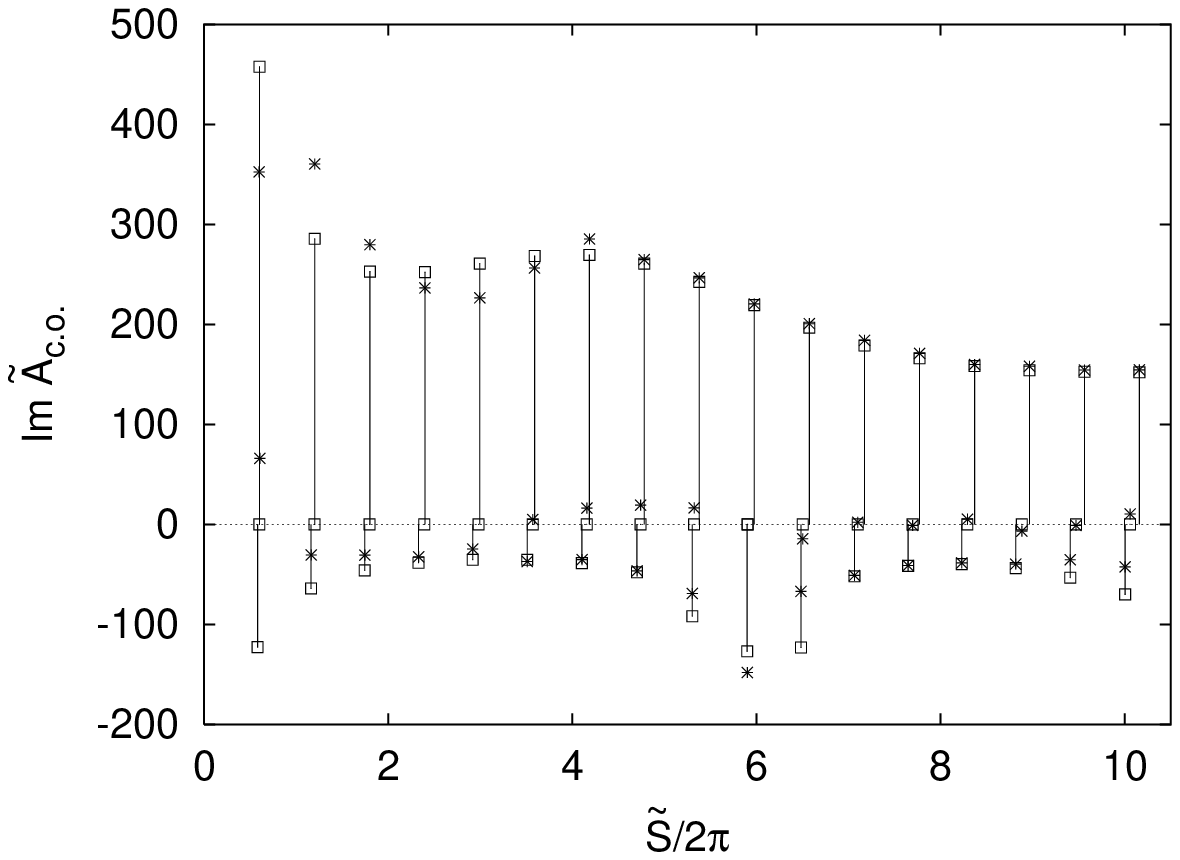}
  \caption{High-resolution recurrence spectrum for $\tilde E=-1.4$ and
  $\tilde F=0.1$. Sticks and squares: semiclassical closed-orbit
  amplitudes, stars: harmonic inversion of the quantum spectrum.}
  \label{RecurReImFig}
\end{figure}

For low scaled actions, where only few closed orbits exist, the
high-resolution analysis can be applied. Results are shown in
figure~\ref{RecurReImFig}, which compares both the scaled actions and the
real and imaginary parts of the semiclassical amplitudes extracted from the
quantum spectrum to the classical results. For most closed orbits, the
agreement is excellent. Exceptions occur for the shortest orbits, where the
actions of rotator and vibrator orbits are too similar to be
resolved by the harmonic inversion. At somewhat larger actions, the three
orbits in each group fall apart into two rotator orbits with
similar actions and a vibrator orbit with a slightly larger action.

\begin{figure}
  \includegraphics[width=.92\columnwidth]{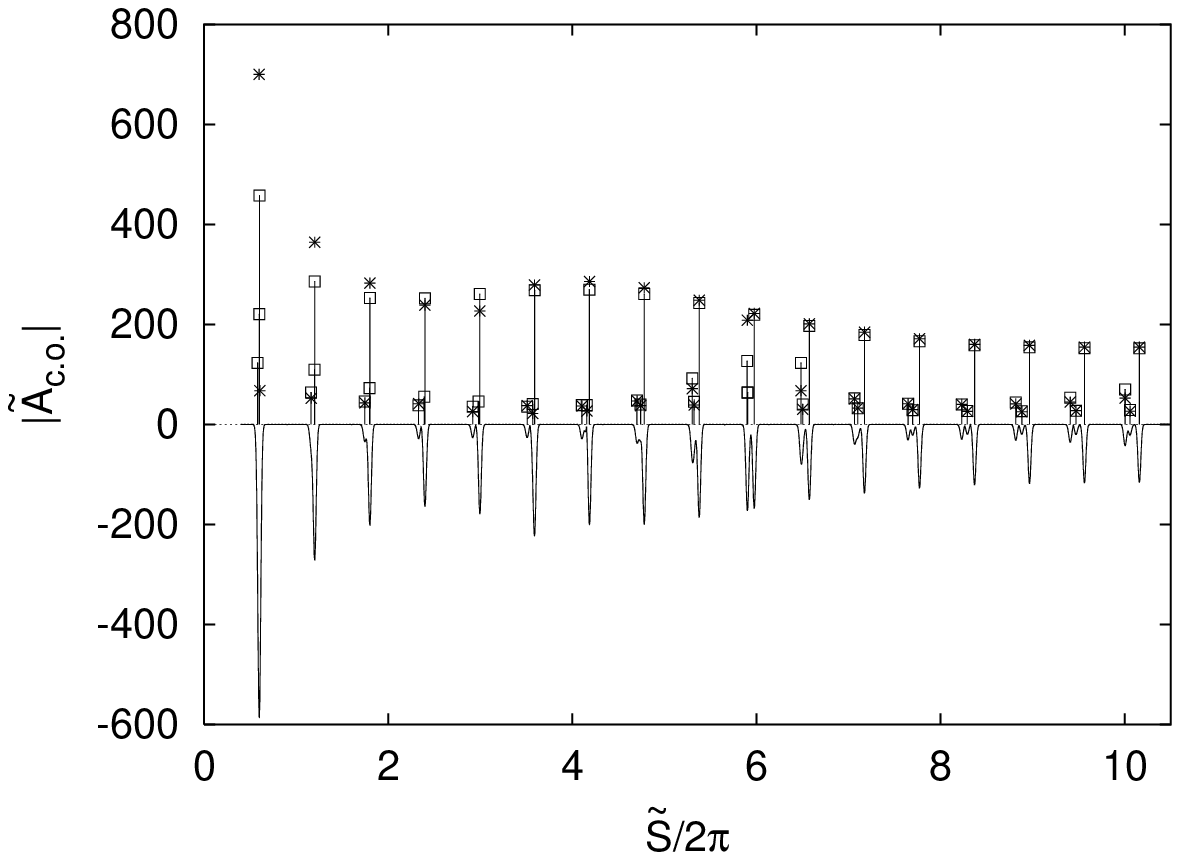}
  \includegraphics[width=.92\columnwidth]{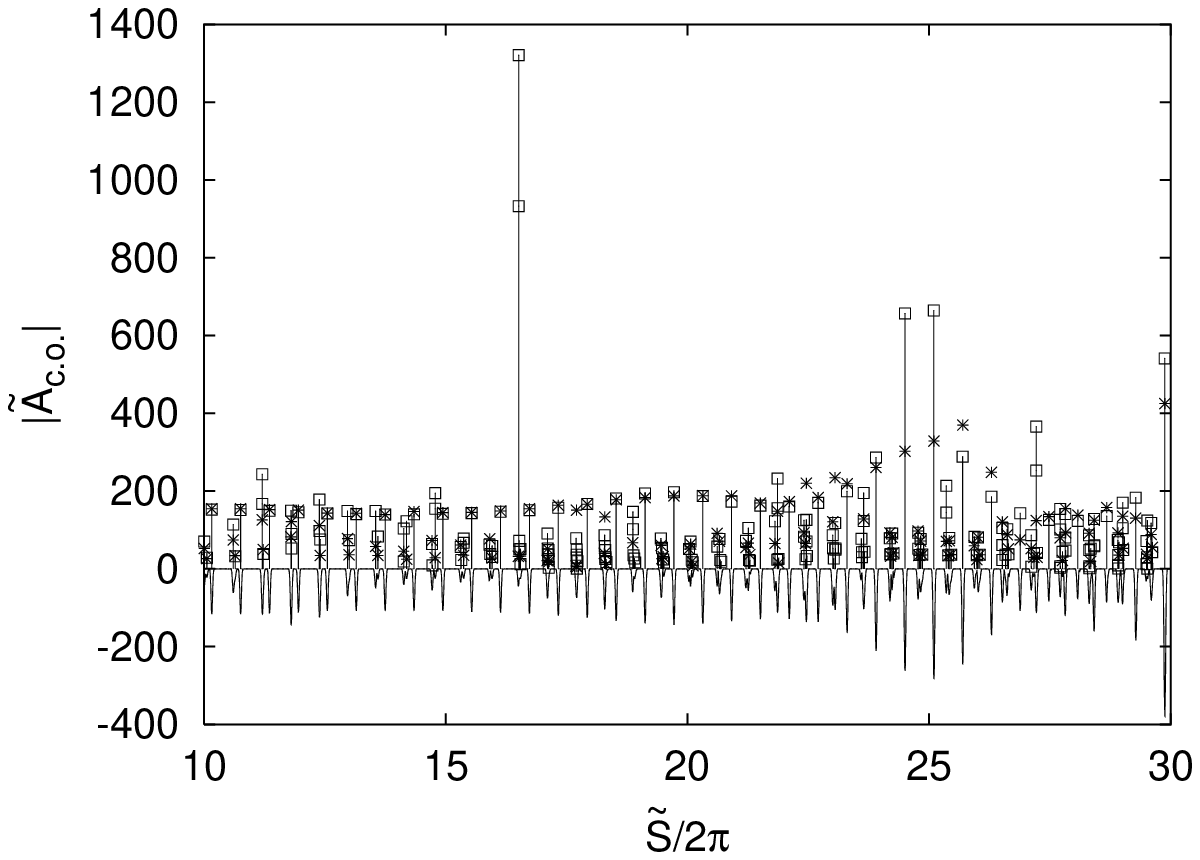}
  \caption{Absolute value of the recurrence spectrum. Sticks and squares: 
  semiclassical closed-orbit amplitudes, stars: harmonic inversion of 
  the quantum spectrum. Solid curve, inverted: Fourier transform (arbitrary
  units).}
  \label{RecurAbsFig}
\end{figure}

These observations can be made even more clearly if the absolute values of
the amplitudes are considered. They are shown in figure~\ref{RecurAbsFig},
where the results of the high-resolution analysis are also compared to
those of the Fourier transform. Notice that for the Fourier transform the
semiclassical amplitude is given by the area under a peak rather than the
peak height, so that an immediate comparison to the high-resolution results
is difficult. In figure~\ref{RecurAbsFig}, the Fourier transform was
arbitrarily scaled so that the peak heights roughly match the values of the
high-resolution amplitudes. For isolated orbits identified both in the
Fourier transform and the high-resolution spectrum, the agreement between
the two methods is excellent. Where several peaks overlap in the
semiclassical spectrum, no direct comparison is possible because the peak
phases cannot be determined from the figure.

Figure~\ref{RecurAbsFig} also extends the results shown in
figure~\ref{RecurReImFig} to higher actions. In this region the
density of closed orbits starts to increase because, on the one hand,
rotators of the first series exist and, on the other, bifurcations of
closed orbits generate additional orbits. Apart from the fact that many
orbits cannot be identified individually even by the high-resolution
method, the most conspicuous feature of figure~\ref{RecurAbsFig} is that
for many orbits the semiclassical amplitudes calculated from the classical
data are considerably larger than those extracted from the quantum
spectrum. In some cases, this is obvious at a glance, but a closer
inspection of the figure reveals that this phenomenon is rather common.
Some specific cases will be described in detail in section~\ref{sec:Uniform}.

The occurrence of exceedingly large semiclassical amplitudes is a
well-known problem of both closed-orbit and periodic-orbit theory. It is
associated with bifurcations of classical orbits, where, in the case of
closed orbits, the stability determinant $M$ vanishes and the closed-orbit
amplitude (\ref{ACrossed}) diverges. Close to the bifurcation, $M$ is
small. The semiclassical amplitude of the bifurcating orbit is therefore
large and exceeds the value determined from the quantum spectrum. In a
classical context, we have shown previously \cite{Bartsch03a} that
vanishing $M$ is both a necessary and sufficient condition for a
bifurcation of closed orbits. In the context of semiclassical closed-orbit
theory, it is necessary to overcome the divergence of the closed-orbit
formula occurring close to a bifurcation. This problem will be addressed in
section~\ref{sec:Uniform}, after the impact of the bifurcations on the
semiclassical signal at hand has been investigated further.

Whereas, in figure~\ref{RecurAbsFig}, the vibrator orbits are sufficiently
isolated to be resolved by both the harmonic inversion and the Fourier
transform across the entire range of actions, the rotators occur in groups
of several orbits having nearly identical actions. They are not resolved
properly by either method. Instead, the Fourier transform produces peaks
describing the collective contribution of the orbits in a group. The
harmonic inversion fits this contribution with fewer actions and amplitudes
than the actual number of orbits. Although the results can be expected to
reproduce the quantum spectrum fairly well, the principal virtue of
the high-resolution analysis -- that it is capable of giving individual
rather than collective contributions -- is lost. It is therefore pointless
to extend the high-resolution analysis to higher actions unless a
significantly longer quantum spectrum can be obtained, and only the Fourier
transform will be used henceforth.

\begin{figure}
  \includegraphics[width=\columnwidth]{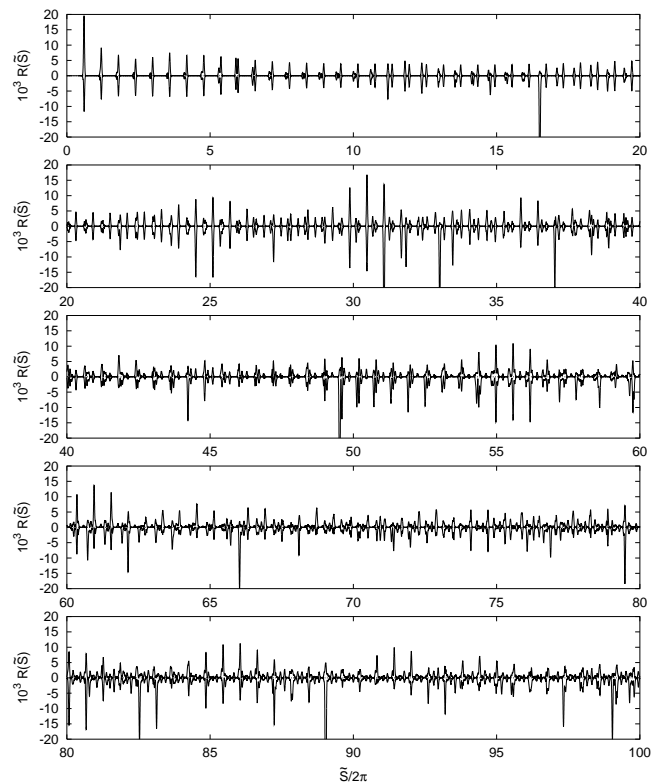}
  \caption{Absolute value $R(\tilde S)$ of the recurrence spectrum with
  $\kappa=10$ (see text). Upper part: Fourier transform of the quantum
  spectrum, lower part (inverted): smoothed semiclassical recurrence
  spectrum.}
  \label{RecurLongFig}
\end{figure}

Figure~\ref{RecurLongFig} displays the Fourier recurrence spectrum with
smoothing $\kappa=10$ for scaled actions up to $\tilde S/2\pi=100$ and
compares it to the semiclassical spectrum. These results extend the
semiclassical analysis of quantum spectra to significantly longer orbits
than investigated in previous studies. They allow a verification of
closed-orbit theory all the way up to the long orbits. It is immediately
apparent from the figure that the quantum recurrence spectrum retains its
pronounced peak structure. This is to be expected from closed-orbit theory,
and indeed the peak locations are given by the actions of closed orbits for
long as well as for short orbits. The basic idea of closed-orbit theory
that recurrence peaks are related to classical closed orbits is therefore
confirmed in principle even for very long orbits.

Even for the largest actions considered, the quantum and semiclassical
recurrence spectra agree quantitatively for some peaks. For most peaks,
however, the peak heights in the quantum and semiclassical spectra
disagree. There are quantum peaks that are smaller in the semiclassical
spectrum or even completely absent. They can be attributed to missing
orbits. On the other hand, in many cases the semiclassical peaks are
significantly higher than the quantum peaks, sometimes by several orders of
magnitude. Exceedingly high peaks can be traced back to bifurcations of
closed orbits if the possibility is ignored that a quantum peak can be
small because orbits missing in the semiclassical spectrum interfere
destructively with the orbits present. This latter mechanism becomes the
more implausible the larger the semiclassical peak is in comparison to the
quantum peak.

Taken together, the effects of missing orbits and of bifurcating orbits
distort the semiclassical recurrence spectrum to the point where it can no
longer be expected to provide a suitable basis for a quantization.  A close
inspection of the recurrence spectrum suggests that the problem posed by
bifurcating orbits is more severe. Exceedingly high peaks do not only occur
frequently, but in addition the very fact that they are high increases
their detrimental effect on the semiclassical photo-absorption
spectrum. Unless a suitable scheme for dealing with
bifurcating orbits can be devised, no improvement of the semiclassical
signal can be expected. We therefore turn to a description of the
semiclassical treatment of bifurcating orbits by means of uniform
approximations.

\section{Uniform approximations}
\label{sec:Uniform}

\subsection{The construction of uniform approximations}
\label{ssec:UnifConst}

Exceedingly large contributions of single orbits to a semiclassical
spectrum arise when the orbits are too close to a bifurcation to be
regarded as isolated, as is implicitly assumed by the stationary-phase
approximation used in the derivation of the closed-orbit formula. Uniform
approximations furnish a collective contribution of all orbits involved in
a bifurcation. This solution was first suggested by Ozorio de Almeida and
Hannay \cite{Ozorio87} in the context of periodic-orbit theory. Their
original approach was extended by different authors
\cite{Sieber96,Schomerus97b,Sieber98,Main97a}, so that today uniform
approximations are a well-established tool of semiclassical physics.
In reference~\cite{Bartsch03a}, we identified two types of generic
closed-orbit bifurcations of codimension one. The pertinent uniform
semiclassical approximations will be derived in what follows.

In most cases of interest, a bifurcation destroys real orbits and turns
them into complex ghost orbits that exist in the complexified classical
phase space. Ghost orbits can yield palpable contributions to semiclassical
spectra \cite{Kus93,Main97a}. In particular, their knowledge is essential
for the construction of uniform approximation. For the generic closed-orbit
bifurcations, the ghost orbits were described along with the real orbits in
reference~\cite{Bartsch03a}.

Of particular importance is the observation that bifurcations of
codimension higher than one are relevant to semiclassics, although on a
classical level they are not generically encountered. They appear as
sequences of generic bifurcations, which, if the individual bifurcations
are sufficiently close, must be described collectively by a single uniform
approximation. Several ex\-amples of uniform approximations for these
complicated bifurcation scenarios have been described in the literature
\cite{Main97a,Schomerus97a,Schomerus98,Bartsch99a}.

The principal requirement a uniform approximation must satisfy is to
asym\-pto\-ti\-cal\-ly reproduce the known isolated-orbits approximation
when the distance from the bifurcation grows large, because in this limit
the stationary-phase approximation can be expected to be accurate. In the
following, we will advocate a somewhat heuristic technique for the
construction of a uniform approximation, which is easy to handle and yields
a smooth interpolation between the asymptotic isolated-orbits
approximations on either side of the bifurcation. It will first be
described in general terms. Subsequently, uniform approximations
describing the generic types of codimension-one bifurcations of closed
orbits will be derived.

A bifurcation scenario is
described by a normal form $\Phi_a(t)$ depending on $n\ge 1$ variables~$t$
and $m\ge 1$ parameters~$a$ such that for any fixed value of the
parameters $a$ there are stationary points of $\Phi_a(t)$ corresponding to
the closed orbits involved in the bifurcation. The parameters $a$ must then
depend on the energy $E$ to reproduce the bifurcations of the closed orbits.

For the uniform approximation we make the ansatz
\begin{equation}
  \label{UnifAnsatz}
  \Psi(E) = I(a) \,\re^{\ri S_0(E)}
\end{equation}
with
\begin{equation}
  \label{UnifI}
  I(a)=\int_{\mathbb{R}^n} d^nt\,p(t)\,\re^{\ri \Phi_a(t)} \;.
\end{equation}
Here, the functions $S_0(E)$ and $p(t)$ as well as the parameter values
$a(E)$ have to be determined. All of them must be smooth functions of $E$.

To find the asymptotic behavior of the uniform approximation
(\ref{UnifAnsatz}) far from the bifurcations, (\ref{UnifI}) is evaluated in
the stationary-phase approximation, which yields
\begin{equation}
  \label{UnifSP}
  \Psi(E) \approx 
      \sum_{t_i} \frac{(2\pi\ri)^{n/2}\,p(t_i)}
                      {\sqrt{|\operatorname{Hess} \Phi_a(t_i)|}}\,
                 \re^{\ri (S_0(E)+\Phi_a(t_i))}\,\re^{-\ri\pi\nu_i/2} \;,
\end{equation}
where the sum extends over all stationary points $t_i$ of $\Phi_a(t)$ that
are real at the given $a$, $\operatorname{Hess}\Phi_a$ is the Hessian determinant of
$\Phi_a$, and $\nu_i$ is the number of negative eigenvalues of
$\operatorname{Hess}\Phi_a(t_i)$. This expression is supposed to reproduce the
isolated-orbits approximation
\begin{equation}
  \label{UnifIsol}
  \Psi(E) \approx \sum_{{\rm c.o.\ }i} {\cal A}_i(E)\, \re^{\ri S_i(E)} \;.
\end{equation}
In this case, the sum extends over all closed orbits involved in the
bifurcation that are real at the given energy $E$. If the normal form
$\Phi_a(t)$ has been chosen suitably, there is a one-to-one correspondence
between these orbits and the stationary points $t_i$. A comparison
of~(\ref{UnifSP}) to~(\ref{UnifIsol}) yields the conditions
\begin{equation}
  \label{UnifAction}
  S_i(E) = S_0(E) + \Phi_a(t_i)
\end{equation}
and
\begin{equation}
  \label{UnifAmp}
  {\cal A}_i(E) = \frac{(2\pi\ri)^{n/2}\,p(t_i)}
                       {\sqrt{|\operatorname{Hess} \Phi_a(t_i)|}}\,
                  \re^{-\ri\pi\nu_i/2} \;.
\end{equation}
These equations must be valid for real orbits.  In most bifurcation
scenarios, all orbits are real at least at certain energies. In these
cases, it appears natural to postulate (\ref{UnifAmp}) also to hold for
ghost orbits. The parameter values one obtains are then smooth functions of
the energy even at the bifurcations where the orbits become ghosts. In some
instances, bifurcations involving only ghost orbits occur
\cite{Bartsch99a,Bartsch99b}. In these cases, the condition~(\ref{UnifAmp})
still produces smoothly varying parameters and enforces the desired
asymptotics.

The numbers~$\nu_i$ of negative eigenvalues change discontinuously at a
bifurcation. For orbits which are real on either side of the bifurcation,
so do the Maslov indices contained in the semiclassical amplitudes~${\cal
A}_i$. These changes must compensate each other if the values~$p(t_i)$ are
to be continuous across the bifurcation. For these orbits, therefore, the
change of Maslov index occurring in a bifurcation must be equal to the
change in $\nu_i$ and can be determined from the normal form. For ghost
orbits, Maslov indices are not well defined classically. They must be
chosen such as to make $p(t_i)$ continuous.

The normal form parameters~$a$ and the action $S_0(E)$ can be determined
from~(\ref{UnifAction}). They usually turn out to be unique. The amplitude
function~$p(t)$, on the contrary, is unknown. Once the parameters~$a$
have been found, (\ref{UnifAmp}) specifies its values $p(t_i)$ at the
stationary points of $\Phi_a(t)$. These values, of course, do not suffice
to identify $p(t)$ uniquely, so that there is considerable freedom in the
choice of $p(t)$.
Usually, if there are $k$~orbits participating in the bifurcation scenario,
we will approximate $p(t)$ by a polynomial of degree~$k-1$. This choice is
justified by the observation that the uniform approximation is needed only
close to a bifurcation, where all orbits are close to $t=0$. Thus, in the spirit of the stationary-phase approximation, the
dominant contributions to the integral~(\ref{UnifI}) stem from the
neighborhood of~$t=0$, whereas the regions of large~$t$ do not
contribute. A suitable approximation to~$p(t)$ must therefore be precise
close to the origin. This is achieved by a Taylor series expansion, which
leads to the polynomial ansatz.

Simple as it might appear, however, this choice can bring about a
mathematical difficulty: A polynomial~$p(t)$ diverges as $t\to\infty$, so
that there is no guarantee that the integral~(\ref{UnifI}) will
converge. If it does not, its divergence is an artefact of the choice of
$p(t)$, because by construction the regions of large $t$ should not
significantly influence the value of the integral. In this case, a suitable
regularization scheme must be applied. It can be justified by verifying
that the regularized integral possesses the correct asymptotics.

A slightly simpler form of the uniform approximation is obtained if the
function $p(t)$ is assumed to be a constant. This approximation does not
exactly reproduce the desired asymptotics, but as the transition across the
bifurcation mainly results in a change of the stationary points of
$\Phi_a(t)$ rather than essential changes in $p(t)$, it can be expected to
capture the principal features.

It is clear from the above description that there is a certain
arbitrariness in the procedure. This arbitrariness can be reduced to the
choice of a suitable amplitude function~$p(t)$, because by the splitting
lemma and the classification theorems of catastrophe theory
\cite{Castrigiano93}
the uniform approximation can always be brought into the
form~(\ref{UnifAnsatz}) by a suitable coordinate transformation, provided a
normal form $\Phi_a(t)$ equivalent to the actual action function is
given. 

In the following sections, uniform approximations for the two generic
co\-dimen\-sion-one bifurcations described in \cite{Bartsch03a} will be
derived along the lines given here. They turn out to be analogous to those
for isochronous and period-doubling bifurcations of periodic orbits given
by Schomerus and Sieber \cite{Schomerus97b}.

\subsection{The fold catastrophe uniform approximation}
\label{ssec:UnifFold}

The simplest closed-orbit bifurcation is the creation of two orbits in a
tangent bifurcation. It is described by the fold catastrophe
\begin{equation}
  \label{Fold}
  \Phi_a(t) = \frac 13 t^3 - a t \;.
\end{equation}
This normal form has stationary points at $t=\pm\sqrt a$, which are real if
$a>0$. Its stationary values are~(\ref{FoldVal})
\begin{equation}
  \label{FoldVal}
  \Phi(\pm\sqrt a) = \mp \frac 2 3\, a^{3/2} \;.
\end{equation}
By~(\ref{UnifAction}), the actions $S_1$ and $S_2$ of the bifurcating
orbits must satisfy
\begin{equation}
  \label{FoldAction}
  \begin{split}
    S_1 &= S_0(E) - \frac 2 3\, a^{3/2} \;,\\
    S_2 &= S_0(E) + \frac 2 3\, a^{3/2} \;.
  \end{split}
\end{equation}
For these equations to hold, one must assume $S_1<S_2$ if the orbits are
real and $\text{Im}\,S_1>0$, $\text{Im}\,S_2<0$ if they are ghosts. These
conditions determine how the orbits are to be associated with the
stationary points of $\Phi_a(t)$.  Equation~(\ref{FoldAction}) can be
solved for
\begin{equation}
  S_0(E) = \frac{S_1+S_2}{2}
\end{equation}
and
\begin{equation}
  |a| = \left(\frac 34\,|S_2-S_1|\right)^{2/3} \;.
\end{equation}
The observation that the bifurcating orbits are real if $a>0$ and ghosts if
$a<0$ fixes the sign of $a$. Both $S_0(E)$ and $a$ have thus be determined.

For the semiclassical amplitudes, (\ref{UnifAmp}) yields
\begin{equation}
  \label{FoldAmp}
  \begin{split}
    {\cal A}_1 &=
      \frac{\sqrt{\pi}}{|a|^{1/4}}\,p(+\sqrt{a})\,\re^{+\ri\pi/4}\;,\\
    {\cal A}_2 &=
      \frac{\sqrt{\pi}}{|a|^{1/4}}\,p(-\sqrt{a})\,\re^{-\ri\pi/4}\;.\\
  \end{split}
\end{equation}
With the ansatz
\begin{equation}
  p(t) = \frac{p_0}{2\pi} + \frac{p_1}{2\pi}\,t
\end{equation}
for the amplitude function $p(t)$, we can solve for the parameters $p_0$
and $p_1$ to obtain
\begin{equation}
  \begin{split}
    p_0 &= \sqrt{\pi}\,|a|^{1/4} \,\re^{-\ri\pi/4}
               ({\cal A}_1 + \ri {\cal A}_2) \;,\\
    p_1 &= \sqrt{\pi}\,\frac{|a|^{1/4}}{\sqrt a}\, \re^{-\ri\pi/4}
               ({\cal A}_1 - \ri {\cal A}_2) \;.\\
  \end{split}
\end{equation}
The uniform approximation thus takes the form
\begin{equation}
  \label{UnifFold}
  \Psi(E) = (p_0 I_0 + p_1 I_1)\re^{\ri S_0(E)}
\end{equation}
with
\begin{equation}
  I_k = \frac{1}{2\pi} \int dt\, t^k \,\re^{\ri\Phi_a(t)} \;.
\end{equation}
The integral $I_0$ can be evaluated in terms of the Airy function
\cite{Abramowitz} as
\begin{equation}
  I_0 = \operatorname{Ai}(-a) \;,
\end{equation}
whereas $I_1$ is given by its derivative
\begin{equation}
  I_1 = \ri\frac{d}{da} I_0 = -\ri\operatorname{Ai}'(-a) \;.
\end{equation}
With these results, the uniform approximation (\ref{UnifFold}) can be
computed once the classical quantities $S_1, S_2$ and ${\cal A}_1,{\cal
A}_2$ are known. After some rearrangements, (\ref{UnifFold}) can be found
to agree with the uniform approximation derived by Schomerus and Sieber
\cite{Schomerus97b} for isochronous bifurcations  of periodic orbits,
although its present form is much simpler.

\subsection{The cusp catastrophe uniform approximation}
\label{ssec:UnifCusp}

The normal form for the symmetrized cusp catastrophe is given
by
\begin{equation}
  \label{Cusp}
  \Phi_a(t) = \frac 14 t^4 - \frac 12 a t^2 \;.
\end{equation}
It has stationary points at $t=0$ and $t=\pm\sqrt a$ and describes a
pitchfork bifurcation, where two asymmetric orbits bifurcate off an orbit
invariant under a reflection. We denote their actions and amplitudes by
$S_{\rm sym}, S_{\rm asym}$ and ${\cal A}_{\rm sym}, {\cal A}_{\rm asym}$,
respectively, where ${\cal A}_{\rm asym}$ is understood to be the
cumulative amplitude of both asymmetric orbits.

As $\Phi_a(t=0)=0$,  the reference action $S_0(E)$ must be chosen equal to
the action of the symmetric orbit. The action difference is given by the
stationary value of $\Phi_a(t)$, which is $a^2/4$, so that
\begin{equation}
  \Delta S = S_{\rm sym}-S_{\rm asym} = \frac 14 a^2\;,
\end{equation}
and
\begin{equation}
  a = \pm 2\sqrt{\Delta S} \;.
\end{equation}
The parameter $a$ has to be chosen positive if the asymmetric orbits are
real, and negative otherwise. Here, $\Delta S$ was assumed to be
positive. If it is not, the normal form $\Phi_a(t)$ must be replaced with
$-\Phi_a(t)$, which changes the sign of the stationary values.

Due to the reflection symmetry, the amplitude function must be an even
function of $t$. We make the ansatz
\begin{equation}
  \label{CuspP}
  p(t) = p_0 + p_2 t^2 \;.
\end{equation}
and solve~(\ref{UnifAmp}) for the coefficients
\begin{equation}
  \begin{gathered}
  p_0 = \sqrt{\frac{a}{2\pi}}\,{\cal A}_{\rm sym}\, \re^{\ri\pi/4} \;, \\
  p_2 = \frac{\re^{-\ri\pi/4}}{2\sqrt{\pi a}}
        \left({\cal A}_{\rm asym}-\sqrt 2 \,\ri{\cal A}_{\rm sym}\right) \;.
  \end{gathered}
\end{equation}

The complete uniform approximation reads
\begin{equation}
  \label{UnifCusp}
  \Psi(E) = \int dt\,p(t)\,\re^{\ri\Phi_a(t)} = p_0 I_0 + p_2 I_2
\end{equation}
with
\begin{equation}
  \label{IkCusp}
  I_k = \int dt\, t^k \,\re^{\ri\Phi_a(t)} \;.
\end{equation}
The integral $I_0$ can be evaluated analytically in terms of Bessel
functions \cite{Gradshteyn}:
\begin{equation}
  \begin{split}
  I_0=\frac{\pi}{2}\sqrt{|a|}\, \re^{-\ri a^2/8}
      \Big[&\re^{\ri\pi/8}\,J_{-1/4}\left(\frac{a^2}{8}\right) \\
       &+ \operatorname{sign} a\, \re^{-\ri\pi/8}\,
            J_{1/4}\left(\frac{a^2}{8}\right)
      \Big] \;.
  \end{split}
\end{equation}
Although it is not apparent at first sight, $I_0$ is a smooth function of
$a$. This can be verified if the series expansion \cite{Abramowitz}
\begin{equation}
  J_\nu(x) = \left(\frac x 2\right)^\nu r_\nu(x)
\end{equation}
with $r_\nu(x)$ a power series in $x^2$ is used. In terms of $r_\nu(x)$, 
\begin{equation}
  \begin{split}
  I_0 = \frac{\pi}{2}\re^{-\ri a^2/8}
        \bigg[&2\re^{\ri\pi/8}\,r_{-1/4}\left(\frac{a^2}{8}\right) \\
              &+\frac a2\re^{-\ri\pi/8}\, r_{1/4}\left(\frac{a^2}{8}\right)
        \bigg] \;,
  \end{split}
\end{equation}
which is indeed smooth.
The second integral $I_2$ can be evaluated from
\begin{widetext}
\begin{equation}
  \label{I2Cusp}
  \begin{split}
    I_2 =& \int dt\, 2\ri\,\frac{d}{da}\,\re^{\ri\Phi_a(t)}
        =  2\ri\,\frac{dI_0}{da} \\
        =& \ri\pi\sqrt{|a|}\,\re^{-\ri a^2/8}
           \bigg\{\left(\frac{1}{2a}-\ri\frac{a}{4}\right)
                  \left[\re^{\ri\pi/8}\,J_{-1/4}\left(\frac{a^2}8\right) + 
                        \operatorname{sign} a \,\re^{-\ri\pi/8}
                             J_{1/4}\,\left(\frac{a^2}8\right) \right] \\
           & \phantom{\ri\pi\sqrt{|a|}\re^{-\ri \frac{a^2}8}\bigg\{}+
                  \frac a8 \,\re^{\ri\pi/8}
                       \left[J_{-5/4}\left(\frac{a^2}8\right) - 
                             J_{3/4}\left(\frac{a^2}8\right) \right] \\
           & \phantom{\ri\pi\sqrt{|a|}\re^{-\ri \frac{a^2}8}\bigg\{}
                  +\operatorname{sign} a \,\frac a8 \,\re^{-\ri\pi/8}
                       \left[J_{-3/4}\left(\frac{a^2}8\right) - 
                             J_{5/4}\left(\frac{a^2}8\right) \right]
            \bigg\} \;.
  \end{split}
\end{equation}
\end{widetext}
This derivation contains an interchange of differentiation and integration
which achieves a regularization of the divergent integral $I_2$. It can be
justified by verifying that the asymptotic behavior of (\ref{I2Cusp}) for
$a\to\pm\infty$ agrees with the stationary phase approximation to
(\ref{IkCusp}).

\section{Uniformized recurrence spectra}
\label{ssec:UnifRecur}

The formulae derived in the preceding sections give the uniform
approximations directly in terms of the semiclassical actions and
amplitudes. This circumstance makes them easy to apply to scaled spectra:
we simply put $S=w\tilde S$ and ${\cal A}_{\rm c.o.} = w^{-1} {\cal
\widetilde A}_{\rm c.o.}$. As $w$ is varied, the bifurcation is not encountered
because the classical mechanics does not change, so that the
isolated-orbits approximation does not actually diverge. However, if $w$ is
small, the action differences between the bifurcating orbits are also
small, so that the presence of the bifurcation is felt and the
isolated-orbits formula produces exceedingly large contributions. For large
$w$, the action differences also grow large, so that the isolated-orbits
approximation should be recovered in the limit of large $w$.

\begin{figure}
  \includegraphics[width=\columnwidth]{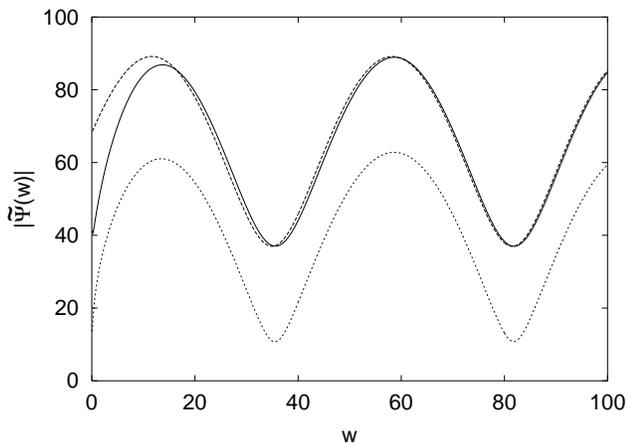}
  \caption{Uniform approximation (\ref{UnifCusp}) for the
  scaled spectrum at $\tilde E=-1.4$ and $\tilde F=0.2$.}
  \label{UnifScalFig}
\end{figure}

These findings are illustrated in figure~\ref{UnifScalFig} for a pitchfork
bifurcation taking place in the first series of rotators at a repetition
number $\mu=57$. At $\tilde E=-1.4$, the bifurcation takes place at the
scaled electric field strength $\tilde F=0.09014$. The data shown in
figure~\ref{UnifScalFig} was calculated for $\tilde E=-1.4$ and $\tilde
F=0.2$, which is sufficiently far away from the bifurcation for the
asymptotic regieme to be reached within the range of $w$ shown. As
anticipated, in the limit of $w\to\infty$ the complete uniform
approximation agrees with the isolated-orbits formula. The simple
approximation also reproduces the beats correctly, but it has a smaller
amplitude.

\begin{figure}
  \includegraphics[width=\columnwidth]{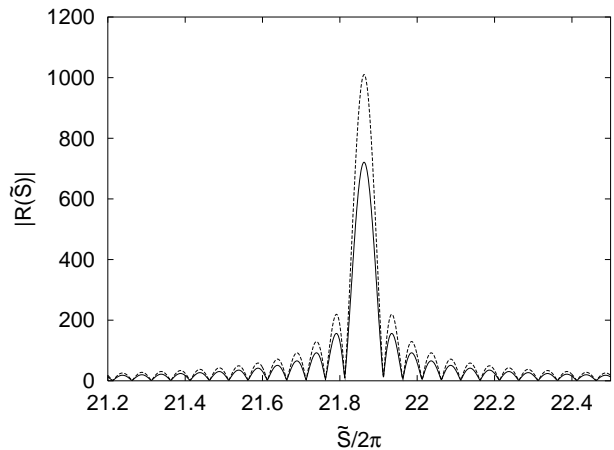}
  \caption{Contribution to the recurrence peak calculated from the uniform
  approximation (solid line) and the isolated-orbits approximation (dashed
  line) for the same bifurcation as in figure~\ref{UnifScalFig}, $\tilde
  E=-1.4$ and $\tilde F=0.1$.}
  \label{UnifRecFig}
\end{figure}
 
The scaled uniform approximation can be used to improve the semiclassical
recurrence spectrum, but this requires some effort: Whereas the
isolated-orbits approximation yields $\delta$ function peaks in the
recurrence spectrum, which are replaced with Gaussians due to the smoothing
of the recurrence spectrum (see section~\ref{sec:SclRec}), the uniform
approximation is a complicated function of $w$. It must be subjected to a
numerical Fourier transform in the same was as the quantum spectrum if its
contribution to the recurrence spectrum is to be evaluated. Because a
bifurcation involves orbits with roughly equal actions, the uniform
approximation will produce a recurrence peak at the appropriate action. An
example is shown in figure~\ref{UnifRecFig}. It was calculated for the
bifurcation already described in figure~\ref{UnifScalFig}. The Gaussian
smoothing used in section~\ref{sec:SclRec} was replaced with a rectangular
window, so that a number of side peaks appear. In this case, the Fourier
transform of both the uniform approximation and the isolated-orbits
approximation was taken over the rectangular window $w\in [40,60]$. The
bifurcating orbits have the scaled action $\tilde S/2\pi\approx 21.86$,
which is where the Fourier peaks are centered in both approximations. The
peak produced by the uniform approximation is considerably smaller.

If this uniformization procedure is carried out for all excessively high
bifurcation peaks, it should be possible to bring the semiclassical
recurrence spectrum in figure~\ref{RecurLongFig} into agreement with its
quantum counterpart. In practice, however, several obstacles stand in the
way. First of all, in many cases ghost orbits must be included in the
uniform approximation. They must be found and identified as pertinent to a
given bifurcation before the uniformization can be performed. Furthermore,
even if all relevant orbits are real, those orbits connected with each
other in a bifurcation must be recognized in the data set. This is by no
means an easy task. For example, if in a given series of rotators and for a
given winding number a quartet of orbits appears, there are two different
doublet orbits out of which they may have bifurcated, and it is not clear
in general which of them must be taken for the uniform approximation. In a
single case, this can be found out fairly comfortably by hand. If many
orbits are to be classified, however, it is essential to do the grouping
automatically. We have not yet been able to devise a practical algorithm
for this task, so that an automatized uniformization of all bifurcation
peaks is presently impossible.

Apart from these rather technical difficulties, there are also some
obstacles of more fundamental importance. Consider, e.g., the two high
semiclassical peaks at $\tilde S/2\pi\approx 25$ in
figure~\ref{RecurAbsFig}. They are notably too high, and they are
well-isolated from neighboring recurrence peaks, so that they may appear
to be the ideal testing ground for the uniformization procedure. These
peaks are generated by vibrators with repetition numbers $\mu=41$ and
$\mu=42$, respectively. The pertinent bifurcation scenarios were described
in figures 17 and 18 of \cite{Bartsch03a}. The ``simple'' scenario taking
place at $\mu=41$ consists of two orbits being generated in the rotational
symmetry-breaking at $\tilde F=0$, followed by a tangent bifurcation
destroying one of them and a third orbit. To smooth this bifurcation peak,
a uniform approximation describing the complete scenario must be found,
which requires the construction of a suitable normal form. Although a
uniform ap\-proxi\-ma\-tion for the symmetry-breaking is available
\cite{Tomsovic95,Ullmo96,Sieber97}, the derivation of the pertinent normal
form relies on principles different from the catastrophe theory
classification used here, and it is not clear how these two can be united
into a single normal form. Thus, the construction of a uniform
approximation for this bifurcation scenario, and even more so for the more
complicated scenario at $\mu=42$, remains an open problem to be solved in
the future. It can be solved within the framework of uniformization
presented in section~\ref{ssec:UnifConst}, but will require a novel way of
constructing normal forms.

The approach to high-resolution semiclassical quantization relies on the
harmonic inversion of a Fourier transformed semiclassical spectrum, i.e. of
a recurrence spectrum. The above method of uniformizing the
bifurcation-induced excessively high recurrence peaks in a semiclassical
spectrum would therefore, if it could be implemented systematically, also
pave the way for the inclusion of uniform approximations into a
high-resolution semiclassical quantization, which has not been possible so
far. We were able to demonstrate the feasibility of our method by way of
example for the hydrogen atom in an electric field \cite{Bartsch02a}, which
is less demanding classically. Its application to the crossed-fields
hydrogen atom, however, remains open for future work.

\section{Conclusion}
\label{sec:Conc}

For the first time, a high-resolution semiclassical quantization of the
hydrogen atom in crossed electric and magnetic fields has been
presented. It achieved the identification of the strong spectral lines in
different $n$-manifolds. By means of a detailed semiclassical analysis of
the pertinent quantum spectrum, it was shown that bifurcations of closed
orbits play a crucial role in the semiclassical spectrum and preclude the
resolution of finer details in the semiclassical spectrum. They pose a
particular challenge to the semiclassical quantization because they require
a special treatment by uniform approximations.

A simple heuristic scheme for the construction of uniform approximations
has been proposed. Its simplicity and efficacy was demonstrated by a
derivation of the uniform approximations for the codimension-one generic
bifurcations of closed orbits.

We have devised a general method for the inclusion of uniform
approximations in a high-resolution semiclassical quantization by harmonic
inversion. In a recent publication \cite{Bartsch02a} it was successfully
applied to the hydrogen atom in an electric field. In the case of the
crossed-fields hydrogen atom, the diversity and complexity of the
bifurcation scenarios encountered so far hinders the systematic
implementation of the uniformization procedure. The treatment of all
relevant bifurcations and the calculation of a detailed semiclassical
spectrum thus remain challenging tasks for future studies.


\end{document}